\documentclass[preprint,prd,nofootinbib,onecolumn,showkeys,showpacs]{revtex4}%
\usepackage{amsfonts}
\usepackage{amsmath}
\usepackage{amssymb}
\usepackage{graphicx}
\usepackage{rotating}
\usepackage{float}%
\providecommand{\U}[1]{\protect\rule{.1in}{.1in}}

\ifx\pdfoutput\relax\let\pdfoutput=\undefined\fi
\newcount\msipdfoutput
\ifx\pdfoutput\undefined\else
\ifcase\pdfoutput\else
\ifx\paperwidth\undefined\else
\ifdim\paperheight=0pt\relax\else\pdfpageheight\paperheight\fi
\ifdim\paperwidth=0pt\relax\else\pdfpagewidth\paperwidth\fi
\fi\fi\fi
\begin{document}
\preprint{ }
\title[Kinematical Conformal Cosmology - Part II]{Kinematical Conformal Cosmology: fundamental parameters from astrophysical observations}
\author{Gabriele U. Varieschi}
\affiliation{Department of Physics, Loyola Marymount University - Los Angeles, CA 90045,
USA\footnote{Email: gvarieschi@lmu.edu}}

\begin{abstract}
We continue our presentation of an alternative cosmology based on conformal
gravity, following our kinematical approach introduced in a recent paper. In
line with the assumptions of our model, which proposes a closed-form
expression for the cosmic scale factor $R(t)$, we first revise the Hubble and
deceleration parameters and also introduce modified cosmological distances,
analyzing in particular the case of the luminosity distance.

Our kinematical conformal cosmology is then able to explain the anomalous
acceleration of the Pioneer spacecraft, as due to a local region of
gravitational blueshift. From the reported values of the Pioneer anomaly we
also compute the current value of our first fundamental parameter, $\gamma
_{0}=1.94\times10^{-28}\ cm^{-1}$, in line with the original estimate by P.
Mannheim of this quantity.

Our second fundamental parameter, $\delta_{0}=3.83\times10^{-5}$, interpreted
as the current value of a cosmological time variable, is derived from a
detailed fitting of type Ia Supernovae \textquotedblleft
gold-silver\textquotedblright\ data, producing Hubble plots of the same
quality of those obtained by standard cosmology, but without requiring any
dark matter or dark energy contribution.

If further experiments will confirm the presence of an anomalous frequency
blueshift in the outer region of the Solar System, as described by our model,
kinematical conformal cosmology might become a viable alternative to standard
cosmological theories.

\end{abstract}
\startpage{1}
\endpage{ }
\maketitle
\tableofcontents

\section{\label{sect:introduction}Introduction}

This paper is the second part of a project aimed at introducing an alternative
cosmology based on Conformal Gravity (CG), as originally proposed by H. Weyl
(\cite{Weyl:1918aa}, \cite{Weyl:1918ib}, \cite{Weyl:1919fi}) and recently
revisited by P. Mannheim and D. Kazanas (\cite{Mannheim:1988dj},
\cite{Kazanas:1988qa}, \cite{Mannheim:2005bf}). In the first paper on the
subject \cite{Varieschi:2008fc} (paper I, in the following) we presented the
mathematical foundations of our new kinematical approach to Conformal
Cosmology. This was based on a critical re-analysis of fundamental
astrophysical observations, starting with the cosmological redshift, and on
the fact that modern metrology defines our common units of length and time
using non-gravitational physics, i.e., through emission/propagation/absorption
of electromagnetic waves or similar phenomena.

Since the laws of electromagnetism are notoriously invariant under a conformal
transformation, we argued that on a cosmological scale a conformal
\textquotedblleft stretching\textquotedblright\ of the space-time might be
present and might yield to an effective change in wavelength/frequency of
electromagnetic radiation, equivalent to the observed cosmological redshift
(or blueshift, if any). As the origin of this presumed conformal stretching of
the metric in the Universe can only be gravitational, we searched its
connection with existing theories of gravity that allow this possible
conformal symmetry. Our attention was focused on Weyl's Conformal Gravity,
since this is the simplest known conformal generalization of Einstein's
General Relativity (GR). Weyl's theory is also based on the same principles
and assumptions of GR, such as the equivalence principle and other
foundational concepts.

The complexity of CG, in particular its fourth order field equations, as
opposed to Einstein's second order equations, have rendered this theory quite
intractable until Mannheim-Kazanas (MK) found the first complete solutions,
such as the exterior solution for a static, spherically symmetric source
(\cite{Mannheim:1988dj}, \cite{Kazanas:1988qa}) and they also showed that it
reduces to the classic Schwartzschild solution in the limit of no conformal
stretching. In addition, the MK solution is able to interpolate smoothly
between the classic Schwartzschild solution and the Robertson-Walker (RW)
metric, through a series of coordinate transformations based on the conformal
structure of the theory.

It was precisely this ability to transform from the conformal MK\ solution to
the standard RW metric, used to describe the cosmological expansion, which
convinced us that a universal conformal stretching might be able to mimic the
expansion of the Universe. The gravitational origin of this conformal
stretching should also lead to the change of observed wavelength/frequency of
cosmic radiation. The principles of General Relativity, which still apply to
its conformal extension, naturally propose such a mechanism: the gravitational redshift.

We have shown in our first paper how the original MK potential can support
this explanation and how the chain of transformations, from Static Standard
Coordinates used in the MK solution to the Robertson Walker coordinates, can
lead to a unique expression of the cosmic scale factor $R$. In this way, the
conformal symmetry of the Universe is \textquotedblleft
kinematically\textquotedblright\ broken and the precise amount of stretching
at each space-time point can be determined, once certain parameters of the
original MK potential are measured.

In this second paper we will show how we can determine these fundamental
parameters using astrophysical data, such as the luminosities of type Ia
Supernovae (SNe Ia) and others. In this way our kinematical conformal
cosmology might become a viable alternative model for the description of the
Universe, with the advantage of avoiding most of the controversial features of
the standard model, such as dark energy, dark matter, inflationary phases, etc.

In the next section we will review our cosmological solutions from paper I,
then we will obtain expressions for the Hubble constant and deceleration
parameter and also revise the definitions of the standard cosmological
distances. In Sect. \ref{sect:cosmological_parameters} we will fit current
astrophysical data in order to compute our cosmological parameters and check
the consistency of our model. Finally, in Sect. \ref{sect:consequences}, we
will explore the immediate consequences of our model, in terms of the behavior
of fundamental constants and other physical quantities.

\section{\label{sect:kinematical_conformal_cosmology}Kinematical Conformal
Cosmology}

\subsection{\label{sect:summary}Summary of results from paper I}

In our first paper \cite{Varieschi:2008fc}, we have essentially worked with
two sets of space-time coordinates. We started with Static Standard
Coordinates - SSC ($r,t,\theta,\phi$) which are used to express the MK
solution for a static, spherically symmetric source, and then we have seen
how, far away from massive sources, the MK\ metric can be transformed into the
RW one, by employing a new set of space-time coordinates, denoted in bold type
($\mathbf{r},\mathbf{t},\theta,\phi$), where the angular coordinates are not
affected by the transformations. The cosmic scale factor can be introduced as
a function of both time coordinates as $\mathbf{R}(\mathbf{t})=R(t)/\sqrt
{\left\vert k\right\vert }$, where $k$ is a cosmological parameter, with
dimensions of an inverse square length, originally introduced in the
MK\ solution and whose value we will also determine in this work.

All the space-time coordinates can be turned into dimensionless quantities
($\mathbf{r}$ is already dimensionless) if we use the following definitions:%

\begin{align}
\alpha &  =2\sqrt{\left\vert k\right\vert }r\label{eqn2.1}\\
\chi &  =\sqrt{\left\vert k\right\vert }c(t_{0}-t)\nonumber\\
\zeta &  =\frac{c(\mathbf{t}_{0}-\mathbf{t})}{\mathbf{R}(\mathbf{t}_{0}%
)},\nonumber
\end{align}
where we use a look-back time $(t_{0}-t)$ or $(\mathbf{t}_{0}-\mathbf{t)}$,
since we usually observe radiation emitted in the past at coordinates $(r,t)$
or $(\mathbf{r},\mathbf{t)}$, reaching us at the spatial origin and at our
current time $(r=0,t_{0})$ or $(\mathbf{r=0},\mathbf{t}_{0})$.

In our first paper we used the MK\ metric as a source of a cosmological
gravitational redshift, associated with a redshift parameter $z$ and a cosmic
scale factor $R$, using the SSC $r$ coordinate. By considering null geodesics
and the other coordinate transformations detailed in Ref.
\cite{Varieschi:2008fc}, we were able to write the cosmic scale factor in any
of the variables described above, obtaining the following expressions:%

\begin{align}
1+z  &  =\frac{R(0)}{R(r)}=\left[  1+\delta\alpha-\frac{1}{4}(1-\delta
^{2})\alpha^{2}\right]  ^{-\frac{1}{2}}\label{eqn2.2}\\
1+z  &  =\frac{R(t_{0})}{R(t)}=\cosh\chi-\delta\sinh\chi\nonumber\\
1+z  &  =\frac{\mathbf{R(0)}}{\mathbf{R(r)}}=\sqrt{1+\mathbf{\ r}^{2}}%
-\delta\mathbf{r}\nonumber\\
1+z  &  =\frac{\mathbf{R(t}_{0}\mathbf{)}}{\mathbf{R(t)}}=\left[  \cos\left(
\sqrt{1-\delta^{2}}\zeta\right)  +\frac{\delta}{\sqrt{1-\delta^{2}}}%
\sin\left(  \sqrt{1-\delta^{2}}\zeta\right)  \right]  ^{-1}.\nonumber
\end{align}

To avoid possible misunderstandings, the cosmic scale factor $R$ is considered
a function of the time coordinate ($t$ or the associated $\mathbf{t}$) as in
standard cosmology, but is expressed also as a function of the radial
coordinate $r$ (or $\mathbf{r}$) simply because information from past times is
brought to us by light emitted at those radial positions. We used in paper I a
gravitational redshift mechanism, based on the static MK potential described
in terms of $r$, to explain the cosmological redshift and therefore we have
\textquotedblleft improperly\textquotedblright\ defined the scale factor as a
function of radial coordinates, as shown in the previous equation.

The solutions in Eq. (\ref{eqn2.2}) were obtained for the particular case
$\mathbf{k}=k/\sqrt{\left\vert k\right\vert }=-1$, which was found to be the
only one associated with a possible redshift of gravitational origin (the
other two cases $\mathbf{k}=0,+1$ did not allow for the observed redshift).
The detailed analysis of these solutions can be found again in our paper I.
Here we recall that the solutions in Eq. (\ref{eqn2.2}) are expressed in terms
of another cosmological parameter $\delta$ defined as:%

\begin{equation}
\delta\equiv\frac{\gamma}{2\sqrt{\left\vert k\right\vert }}. \label{eqn2.3}%
\end{equation}
The additional quantity $\gamma$ was also introduced in the original MK
solution and the main objective of this paper is to determine the values of
these three cosmological parameters ($\delta$, $\gamma$ and $k$) linked
together by the previous equation.

Mannheim was able to fit galactic rotation curves without the need of dark
matter and to estimate the current value of $\gamma$ as $\gamma_{Mann}%
=3.06\times10^{-30}%
\operatorname{cm}%
^{-1}$ \cite{Mannheim:1996rv}. In our first paper we argued that the current
value of $\gamma$ is probably close to Mannheim's estimate, but needs to be
computed from more \textquotedblleft local\textquotedblright\ observations. In
addition, $k$ should have a negative value (since $\mathbf{k}=k/\sqrt
{\left\vert k\right\vert }=-1$) while $\delta$ is necessarily limited by
$-1<\delta<+1$, so that its current value is probably small and positive (see
again paper I for details).

Another important hypothesis, introduced in our first paper, is to assume that
$\delta$ (as well as $\gamma$ and $k$) are probably time-varying quantities,
over cosmological ages. In fact, we have proposed that the dimensionless
parameter $\delta$ might constitute an effective cosmological time, varying
from $-1$ to $+1$, so that Eq. (\ref{eqn2.2}) represents the evolution of the
Universe as seen at our \textquotedblleft current time\textquotedblright%
\ $\delta=\delta(t_{0})$. The most general description is obtained by letting
$\delta$ vary in the allowed interval, in all the preceding formulas. If this
interpretation is correct it is possible to write the scale factor directly as
a function of the variable $\delta$ and of its current value $\delta(t_{0})$
as follows:%

\begin{equation}
{\small 1+z=}\frac{R\left[  \delta(t_{0})\right]  }{R(\delta)}=\sqrt
{\frac{1-\delta^{2}(t_{0})}{1-\delta^{2}}}, \label{eqn2.4}%
\end{equation}
a simple \textquotedblleft semi-circular\textquotedblright\ evolution
illustrated in Fig. 2 of paper I. The complete connections between all these
variables are also discussed in details and summarized in Table 1 of our first paper.

The next step is to check our model against current astrophysical data, in
order to establish it as a viable alternative to current cosmology. The
cosmological parameters introduced above also need to be evaluated and
connected to standard cosmological quantities such as the Hubble constant and
the deceleration parameter.

\subsection{\label{sect:hubble_constant}The Hubble constant and the
deceleration parameter}

One of the goals of standard cosmology is to determine, both theoretically and
experimentally, the Hubble constant and the deceleration parameter which are
essential to describe the evolution of the Universe. Since in our model the
cosmic scale factor $R$ is determined explicitly by Eq. (\ref{eqn2.2}) it is
not difficult to obtain these important parameters.

We recall that in general the Hubble parameter is defined as $H(t)=\frac
{\overset{\cdot}{R}(t)}{R(t)}$ and the deceleration parameter as
$q(t)=-\frac{\overset{\cdot\cdot}{R}(t)}{R(t)H^{2}(t)}=-\frac{\overset
{\cdot\cdot}{R}(t)R(t)}{\overset{\cdot}{R}{}^{2}(t)}$, with their current-time
values denoted by $H_{0}$ and $q_{0}$. Standard cosmology measurements of the
Hubble constant are usually reported as \cite{Lang:1999ry}:%

\begin{equation}
H_{0}=100\ h\ km\ s^{-1}\ Mpc^{-1}=3.24\times10^{-18}\ h\ s^{-1},
\label{eqn2.5}%
\end{equation}
where $h$ is a number between $0.5$ and $1$.

Following the model discussed in paper I and briefly reviewed above, we can
write $H(t)$ and $q(t)$ by using our fundamental solutions and we can also
express these quantities in terms of either one of the two time coordinates
$t$ or $\mathbf{t}$, introduced previously. As already explained in Sect. IV
of Ref. \cite{Varieschi:2008fc}, our preference goes to the simpler $t$
coordinate, which makes direct contact with our units of time, but we will
also consider the other coordinate $\mathbf{t}$ in the following.

We start by using the SSC time coordinate $t$, which is connected to the
dimensionless look-back time $\chi=\sqrt{\left\vert k\right\vert }c(t_{0}-t)$
as in Eq. (\ref{eqn2.1}). In paper I we have seen in Eqs. (75)-(77) how to
express the first and second order time derivatives of $R$ in terms of $\chi$,
or directly in terms of the redshift parameter $z$. As a consequence, we can
easily write $H(t)$ and $q(t)$ also as a function of $\chi$ or $z$:%

\begin{align}
H(t)  &  =\sqrt{\left\vert k\right\vert }c\left(  \frac{\sinh\chi-\delta
\cosh\chi}{\cosh\chi-\delta\sinh\chi}\right)  =\pm\sqrt{\left\vert
k\right\vert }c\frac{\sqrt{(1+z)^{2}-(1-\delta^{2})}}{(1+z)}\label{eqn2.6}\\
q(t)  &  =\left(  \frac{\cosh\chi-\delta\sinh\chi}{\sinh\chi-\delta\cosh\chi
}\right)  ^{2}-2=\frac{(1+z)^{2}}{(1+z)^{2}-(1-\delta^{2})}-2,\nonumber
\end{align}
where, as in the preceding equations, we use the \textquotedblleft
current\textquotedblright\ value $\delta=\delta(t_{0})$ (or the value at the
time the observations were made). The current-time values of the Hubble and
deceleration parameters are obtained in the limit for $\chi\rightarrow0$ or
$z\rightarrow0$:%

\begin{align}
H(t_{0})  &  =-\frac{\gamma}{2}c\ ;\ H(z=0)=\pm\frac{\gamma}{2}c
\label{eqn2.7}\\
q(t_{0})  &  =q(z=0)=\frac{1}{\delta^{2}}-2.\nonumber
\end{align}

The signs of the quantities in Eqs. (\ref{eqn2.6}) and (\ref{eqn2.7}) can be
explained with the help of the red-solid curve in Fig. 5 of paper I, which
represents the ratio $R(t)/R(t_{0})$. This bell-shaped curve was plotted for a
positive value $\delta=\delta(t_{0})$ and shows a local blueshift area in the
\textquotedblleft past\textquotedblright\ evolution of the Universe, extending
back to a time $t_{rs}$, followed by a redshift region which extends
indefinitely to past times and which should represent the observed
cosmological redshift from past cosmological epochs.

While we will explain the local blueshift region later in this paper (in Sect.
\ref{sect:pioneer}), we simply remark here that this red curve in Fig. 5 is
symmetric around the point of maximum. Therefore, for each value of $z$, we
have two corresponding values of the Hubble parameter (except at the maximum,
for $z_{\min}=\sqrt{1-\delta^{2}}-1$, where $H=0$) and the two related points
on the curve, at the same redshift level, will have equal and opposite
expansion rates. This yields the double sign in the previous expressions for
$H$, when given as a function of $z$.

This applies also to the $z=0$ case, corresponding to the current time $t_{0}$
at which $H(t_{0})=-\frac{\gamma}{2}c$ is negative, but also corresponding to
the time in the past ($t_{rs}$) at which we start observing the cosmological
redshift, with $H(t_{rs})=+\frac{\gamma}{2}c$ a positive quantity. This does
not contradict the current estimates of $H_{0}$ as a positive quantity. They
are based on redshift observations of light coming from galaxies at times in
the past $t\lesssim t_{rs}$, therefore what we call $H_{0}$ in standard
cosmology should be actually indicated as $H(t_{rs})=+\frac{\gamma}{2}c$, a
positive quantity.

On the contrary, in section \ref{sect:pioneer} we will evaluate the current
value of the gamma parameter, by analyzing the local \textit{blueshift} in the
region of our Solar System, corresponding to a negative $H(t_{0}%
)=-\frac{\gamma}{2}c$. In this way we will be able to estimate the gamma
parameter as $\gamma(t_{0})\cong1.94\times10^{-28}\ cm^{-1}$ and this will
allow us to evaluate also the Hubble parameter at the beginning of the
redshift region (time $t_{rs}$), by using the same value of $\gamma$ and the
positive sign in Eq. (\ref{eqn2.7}), or equivalently assuming by symmetry
$\gamma(t_{rs})\cong-1.94\times10^{-28}\ cm^{-1}$ and using $H(t_{rs}%
)=-\frac{\gamma(t_{rs})}{2}c$. Numerically, we estimate:%

\begin{align}
H(t_{0})  &  =-\frac{\gamma(t_{0})}{2}c\cong-2.91\times10^{-18}\ s^{-1}%
\label{eqn2.8}\\
H(t_{rs})  &  =+\frac{\gamma(t_{0})}{2}c=-\frac{\gamma(t_{rs})}{2}%
c\cong+2.91\times10^{-18}\ s^{-1},\nonumber
\end{align}
where again the details of this analysis will be presented later in this paper.

The connection with the standard cosmology value of $H_{0}$ in Eq.
(\ref{eqn2.5}) is immediate: all the standard astrophysical observations which
led to the existing estimates of $H_{0}$ were done by observing celestial
objects in the redshift region,\ therefore for $r\gtrsim r_{rs}$ or $t\lesssim
t_{rs}$, thus%

\begin{align}
H_{0}  &  =H(t_{rs})=+\frac{\gamma(t_{0})}{2}c\cong+2.91\times10^{-18}%
\ s^{-1}=100\ h_{rs}\ km\ s^{-1}\ Mpc^{-1}\label{eqn2.9}\\
h_{rs}  &  \cong0.897\nonumber
\end{align}
in line with all the current estimates of $H_{0}$ \cite{Lang:1999ry}. Our
value, $h_{rs}\cong0.897$, is the direct estimate of the $h$ parameter based
on the Pioneer anomaly data (see Sect. \ref{sect:pioneer}) and is close to
recent determinations by the Wilkinson Microwave Anisotropy Probe (WMAP)
\cite{Spergel:2006hy} $h_{WMAP}\cong0.73\pm0.03$, by the Hubble Space
Telescope Key Project (HST Key Project) \cite{2001ApJ...553...47F}
$h_{HST}\cong0.72\pm0.08$, and others \cite{Riess:2005zi}.\footnote{All these
current determinations of the Hubble constant are obviously based on standard
cosmology and on the current calibration methods of the cosmological distance
ladder. It is beyond the scope of this paper to check these standard
estimates, in view of the changes introduced by our kinematical conformal
cosmology.}

The deceleration parameter at current time $t_{0}$, or at time $t_{rs}$ (in
both cases $z=0$), is given by Eq. (\ref{eqn2.7}) as a function of the
dimensionless $\delta$, which cannot be estimated from the Pioneer data. We
cannot use results for $q$ or similar acceleration parameters coming from
standard cosmology, such as those obtained from type Ia Supernovae analysis,
as they are based on totally different assumptions. We will have to analyze
and re-interpret the concepts of standard candle, luminosity distance, etc.,
before we can estimate $\delta(t_{0})$ and therefore $q(t_{0})$. This will be
done in the following sections.

Before we proceed in this direction, we also want to study the expressions for
the Hubble and deceleration parameters which can be obtained by using other
variables. For example, we can recalculate our expressions using the RW time
variable $\mathbf{t}$ instead, i.e., define the Hubble parameter as
$\mathbf{H(t)=}\frac{\overset{\cdot}{\mathbf{R}}\mathbf{(t)}}{\mathbf{R}%
(\mathbf{t})}$ and the deceleration parameter $\mathbf{q(t)}=-\frac
{\overset{\cdot\cdot}{\mathbf{R}}(\mathbf{t})}{\mathbf{R(t)H}^{2}(\mathbf{t}%
)}=-\frac{\overset{\cdot\cdot}{\mathbf{R}}(\mathbf{t})\mathbf{R(t)}}%
{\overset{\cdot}{\mathbf{R}}{}^{2}(\mathbf{t})}$, where the boldface symbols
denote quantities related to the RW metric, as discussed in the previous
section and in paper I. It is straightforward to obtain expressions similar to
those in Eq. (\ref{eqn2.6}):%

\begin{align}
\mathbf{H}(\mathbf{t})  &  =-\frac{c\sqrt{1-\delta^{2}}}{\mathbf{R(t}%
_{0}\mathbf{)}}\left[  \frac{\delta\cos(\sqrt{1-\delta^{2}}\zeta
)-\sqrt{1-\delta^{2}}\sin(\sqrt{1-\delta^{2}}\zeta)}{\sqrt{1-\delta^{2}}%
\cos(\sqrt{1-\delta^{2}}\zeta)+\delta\sin(\sqrt{1-\delta^{2}}\zeta)}\right]
=\pm\frac{c}{\mathbf{R(t}_{0}\mathbf{)}}\sqrt{(1+z)^{2}-(1-\delta^{2}%
)}\label{eqn2.10}\\
\mathbf{q}(\mathbf{t})  &  =\left[  \frac{\sqrt{1-\delta^{2}}\cos
(\sqrt{1-\delta^{2}}\zeta)+\delta\sin(\sqrt{1-\delta^{2}}\zeta)}{\delta
\cos(\sqrt{1-\delta^{2}}\zeta)-\sqrt{1-\delta^{2}}\sin(\sqrt{1-\delta^{2}%
}\zeta)}\right]  ^{2}=\frac{(1-\delta^{2})}{(1+z)^{2}-(1-\delta^{2}%
)},\nonumber
\end{align}
where again\ $\delta=\delta(t_{0})$. We use here the dimensionless variable
$\zeta=\frac{c(\mathbf{t}_{0}-\mathbf{t})}{\mathbf{R}(\mathbf{t}_{0})}$ or we
express the Hubble and deceleration parameters directly in terms of $z$.

The current-time values of these two parameters are obtained in the limit for
$\zeta\rightarrow0$ or $z\rightarrow0$, respectively:%

\begin{align}
\mathbf{H}(\mathbf{t}_{0})  &  =-\frac{c}{\mathbf{R(t}_{0}\mathbf{)}}%
\delta\ ;\ \mathbf{H}(z=0)=\pm\frac{c}{\mathbf{R(t}_{0}\mathbf{)}}%
\delta\label{eqn2.11}\\
\mathbf{q}(\mathbf{t}_{0})  &  =\mathbf{q}(z=0)=\frac{1}{\delta^{2}%
}-1,\nonumber
\end{align}
with the same interpretation which was given for Eq. (\ref{eqn2.7}). These
expressions for the Hubble and deceleration parameters in the two temporal
variables are obviously connected. It follows from the definitions that
$\mathbf{H}(\mathbf{t})=H(t)\frac{dt}{d\mathbf{t}}$ and $\mathbf{q}%
(\mathbf{t})=q(t)-\frac{d^{2}t/d\mathbf{t}^{2}}{H(t)\left(  dt/d\mathbf{t}%
\right)  ^{2}}$, with the derivatives between time variables given by:%

\begin{align}
\frac{dt}{d\mathbf{t}}  &  =\frac{\sqrt{1-\delta^{2}}}{\sqrt{\left\vert
k\right\vert }\mathbf{R(t}_{0}\mathbf{)}}\frac{1}{\left[  \sqrt{1-\delta^{2}%
}\cos(\sqrt{1-\delta^{2}}\zeta)+\delta\sin(\sqrt{1-\delta^{2}}\zeta)\right]
}=\frac{\left(  1+z\right)  }{\sqrt{\left\vert k\right\vert }\mathbf{R(t}%
_{0}\mathbf{)}}\label{eqn2.12}\\
\frac{d^{2}t}{d\mathbf{t}^{2}}  &  =\frac{c(1-\delta^{2})}{\sqrt{\left\vert
k\right\vert }\mathbf{R}^{2}\mathbf{(t}_{0}\mathbf{)}}\frac{\left[  \delta
\cos(\sqrt{1-\delta^{2}}\zeta)-\sqrt{1-\delta^{2}}\sin(\sqrt{1-\delta^{2}%
}\zeta)\right]  }{\left[  \sqrt{1-\delta^{2}}\cos(\sqrt{1-\delta^{2}}%
\zeta)+\delta\sin(\sqrt{1-\delta^{2}}\zeta)\right]  ^{2}}=\mp\frac
{c(1+z)}{\sqrt{\left\vert k\right\vert }\mathbf{R}^{2}\mathbf{(t}%
_{0}\mathbf{)}}\sqrt{(1+z)^{2}-(1-\delta^{2})},\nonumber
\end{align}
so that the connecting formulas can be easily found.

Similarly, we could also write Hubble and deceleration parameters in terms of
space variables $r$, $\alpha$, or $\mathbf{r}$, since we have in Eq.
(\ref{eqn2.2}) the cosmic scale factor $R$ expressed in all these variables,
but these expressions would not be very useful because the experimental values
of $H_{0}$ and $q_{0}$ are usually referred only to the temporal variables. We
will return in Sect. \ref{sect:pioneer} on the connection with experimental observations.

Finally, from Eq. (\ref{eqn2.4}), it is possible to introduce $H$ and $q$
directly as a function of our cosmological time $\delta$:%

\begin{align}
H(\delta)  &  =\frac{1}{R(\delta)}\frac{dR}{d\delta}=-\frac{\delta}%
{1-\delta^{2}}\label{eqn2.13}\\
q(\delta)  &  =-\frac{\frac{d^{2}R}{d\delta^{2}}R(\delta)}{\left(  \frac
{dR}{d\delta}\right)  ^{2}}=\frac{1}{\delta^{2}}\nonumber
\end{align}
and the expressions in Eqs. (\ref{eqn2.6}) and (\ref{eqn2.10}) will reduce to
those in Eq. (\ref{eqn2.13}), using the transformations outlined in Table 1 of
paper I.

\subsection{\label{sect:luminosity_distance}Luminosity distance and other
cosmological distances}

Before we make contact with experimental data, especially with standard candle
measurements, we need to review the definitions of the cosmological distances,
following the new principles of our kinematical conformal cosmology as
outlined above and in paper I.

Several distances are usually introduced in standard cosmology and used in
conjunction with astronomical observations in order to establish the
\textquotedblleft cosmological distance ladder,\textquotedblright\ i.e., the
different steps employed to measure astronomical distances and the size of the
Universe (for an extensive introduction to the subject see \cite{Webb:1999}
and \cite{Peacock:1999ye}). This process started historically at the time of
Greek astronomy with the determination of the size and radius of our planet
and with the first approximate measurements of the astronomical unit and other
distances within the Solar System.

These estimates, based mainly on parallax methods, were later refined by
modern astronomers and extended to parallax measurements of nearby stars.
These geometrical methods led to the introduction of distances such as the
\textit{parallax distance} $d_{P}$, whose definition is not affected by our
alternative approach to cosmology. As defined in Weinberg's books
(\cite{Weinberg}, \cite{Weinberg2}):%

\begin{equation}
d_{P}\equiv\frac{b}{\theta}=\mathbf{R(t}_{0}\mathbf{)}\frac{\mathbf{r}}%
{\sqrt{1-\mathbf{kr}^{2}}}=\mathbf{R(t}_{0}\mathbf{)}\frac{\mathbf{r}}%
{\sqrt{1+\mathbf{r}^{2}}}, \label{eqn2.14}%
\end{equation}
where $b$ is the impact parameter of light reaching the observer from a
(parallax) distance $d_{P}$ with parallax angle $\theta$, and we have used our
preferred value $\mathbf{k}=-1$ in the right-hand side of the last equation
(see \cite{Lang:1999ry}, \cite{Peacock:1999ye}, \cite{Weinberg} or
\cite{Weinberg2}, for full details on all these distances).

Two more fundamental distances are usually introduced in cosmology, the
\textit{comoving distance} $d_{C}$ (sometimes also called coordinate or
effective distance) and the \textit{proper distance} $d_{prop}$:%

\begin{align}
d_{C}  &  \equiv\mathbf{R(t}_{0}\mathbf{)r}\label{eqn2.15}\\
d_{prop}  &  \equiv\mathbf{R(t)}\int_{\mathbf{0}}^{\mathbf{r}}\frac
{d\mathbf{r}^{\prime}}{\sqrt{1-\mathbf{kr}^{\prime2}}}=\mathbf{R(t)\chi=R(t)}%
\begin{Bmatrix}
\arcsin\mathbf{r\ ;} & \mathbf{k}=+1\\
\mathbf{r\ ;} & \mathbf{k}=0\\
\operatorname{arcsinh}\mathbf{r\ ;} & \mathbf{k}=-1
\end{Bmatrix}
,\nonumber
\end{align}
where $d_{C}$ usually refers to the current-time expansion factor
$\mathbf{R(t}_{0}\mathbf{)}$, while $d_{prop}$ refers to any cosmological time
$\mathbf{t}$ being considered. These definitions are also unchanged in our
cosmology and can be rewritten in terms of other variables using the
transformations in paper I.

Going back to our brief summary of the cosmological distance ladder, modern
parallax techniques can determine star distances up to about $100\ $parsec
($1\ pc=3.086\times10^{18}cm$) and these distance estimates are usually
combined with the measured apparent star luminosities, to obtain their
corresponding absolute luminosities. For this purpose the inverse square law
is typically assumed, introducing the \textit{luminosity distance} $d_{L}$ and
connecting it to the apparent ($l$) and absolute bolometric luminosity ($L$)
of a light source as follows:%

\begin{align}
l  &  =\frac{L}{4\pi d_{L}^{2}}\label{eqn2.16}\\
d_{L}  &  \equiv\sqrt{\frac{L}{4\pi l}}=\frac{\mathbf{R}^{2}(\mathbf{t}_{0}%
)}{\mathbf{R}(\mathbf{t})}\mathbf{r}=(1+z)\mathbf{R}(\mathbf{t}_{0}%
)\mathbf{r.}\nonumber
\end{align}

This definition follows from the consideration that, \textquotedblleft In a
Euclidean space the apparent luminosity of a source at rest at distance $d$
would be $L/4\pi d^{2}$,...\textquotedblright\ (see again Weinberg
\cite{Weinberg}, Sect. 14.4). In the second line of the previous equation, the
factor $\mathbf{R}(\mathbf{t}_{0})/\mathbf{R}(\mathbf{t})=(1+z)$ appears due
to the standard redshift interpretation, namely that photons of energy $h\nu$
are redshifted to energy $h\nu\ \mathbf{R}(\mathbf{t})/\mathbf{R}%
(\mathbf{t}_{0})$ and that the time interval of photon emission $\delta t$ is
also changed to $\delta t\ \mathbf{R}(\mathbf{t}_{0})/\mathbf{R}(\mathbf{t})$.
The total power emitted (energy per unit time) is therefore redshifted by a
combined factor $\mathbf{R}^{2}(\mathbf{t})/\mathbf{R}^{2}(\mathbf{t}%
_{0})=(1+z)^{-2}$ which will enter the denominator of the square root of Eq.
(\ref{eqn2.16}), thus resulting in the final $(1+z)$ factor in the equation
(see \cite{Weinberg} for full details). In other words, the standard redshift
effect is assumed to alter the apparent luminosity of a standard candle placed
far away, so that the $(1+z)$ factor corrects for this effect, while the
absolute luminosity $L$ of our candle is considered fixed and constant. We
will need to change this view in our alternative interpretation.

Before we consider our revised luminosity distance, we recall that
astronomical luminosity measurements are usually expressed in terms of
magnitudes ($m$) using Pogson's law, $m_{1}-m_{2}=-2.5\log_{10}(l_{1}/l_{2})$,
for any two apparent luminosities. Traditionally, the absolute bolometric
luminosity $L$ of a standard candle is defined as the apparent bolometric
luminosity of the same object placed at a reference distance of $10\ pc$, so
that the distance modulus $\mu$ (difference between the apparent and absolute
bolometric magnitudes) will result as follows:%

\begin{equation}
\mu\equiv m-M=5\log_{10}(d_{L}/pc)-5=5\log_{10}(d_{L}/Mpc)+25. \label{eqn2.17}%
\end{equation}

In standard cosmology the luminosity distance on the right-hand side of the
last equation is then expressed as a function of $z$ and other cosmological
parameters, such as the density parameters $\Omega_{M}$ and $\Omega_{\Lambda}%
$, i.e., $d_{L}=d_{L}(z;\Omega_{M},\Omega_{\Lambda})$. The comparison with
experimental observations is usually carried out by fitting the luminosity
distance expression to the measured distance moduli ($\mu$) for several
astrophysical light sources, which can be considered standard candles, i.e.,
assumed having constant intrinsic luminosity. This method has proven to be
particularly reliable when applied to type Ia Supernovae and will be analyzed
in detail in the following sections.

However, our view of the luminosity distance is different from the one
outlined above. First of all, the cosmological redshift is related to the
intrinsic stretching of the space-time fabric at cosmological distances and
over cosmological times. This is realized through the gravitational redshift
mechanism described in Eq. (\ref{eqn2.2}) and not anymore through a
Doppler-like shift in photons energy or change in their emission frequency. In
this view, the correcting factor $(1+z)$ inserted in Eq. (\ref{eqn2.16}) and
described above is no longer necessary: the photons are emitted at the source
with the precise frequency, energy and rate of emission measured by the
Earth's observer (although these differ from the same quantities measured by
an observer near the source).

Therefore, we correct the standard definition of the luminosity distance in
Eq. (\ref{eqn2.16}) by removing the $(1+z)$ factor:%

\begin{equation}
d_{L}\equiv\sqrt{\frac{L_{z}}{4\pi l}}=\mathbf{R}(\mathbf{t}_{0}%
)\mathbf{r=R}(\mathbf{t}_{0})\frac{\delta(1+z)+\sqrt{(1+z)^{2}-(1-\delta^{2}%
)}}{(1-\delta^{2})}, \label{eqn2.18}%
\end{equation}
where we have inserted, on the right-hand side of the equation, our expression
for the coordinate $\mathbf{r}$, introduced in Eq. (71) of our paper I, as a
function of our current value $\delta=\delta(t_{0})$. We also choose the
positive sign in front of the square root to select the solution corresponding
to past redshift, $z>0$ for $\mathbf{r>r}_{rs}=\frac{2\delta}{1-\delta^{2}}$,
which is the correct choice for the subsequent analysis.

In the previous equation we have also indicated a dependence on $z$ of the
absolute luminosity $L_{z}$ of the standard candle being considered (indicated
by the subscript $z$). This is the second fundamental difference in our
analysis of the luminosity distance. We have based our discussion in paper I
on the hypothesis that the fundamental units of measure, such as the meter or
others, depend on the space-time position in the Universe and differ from the
same units of measure at another space-time location. Consequently, we have to
assume that the same hypothesis applies to the luminosity of a
\textquotedblleft standard candle,\textquotedblright\ i.e., we cannot assume
anymore that $L$ is an invariant quantity, when observing these candles placed
at cosmological distances.

In our first paper we have also postulated that space or time intervals are
\textquotedblleft dilated\textquotedblright\ by the $(1+z)$ factor, when
referring to a cosmological location\ characterized by redshift $z$, but we
haven't introduced any similar dependence for the third fundamental mechanical
quantity, i.e., mass. In fact, we have no \textit{a priori} indication of how
masses should scale due to our space-time stretching, therefore we summarize
the scaling properties of the three fundamental quantities as follows:%

\begin{align}
\delta l_{z}  &  =(1+z)\ \delta l_{0}\label{eqn2.19}\\
\delta t_{z}  &  =(1+z)\ \delta t_{0}\nonumber\\
\delta m_{z}  &  =f(1+z)\ \delta m_{0}.\nonumber
\end{align}

The first two lines in this equation simply rewrite Eq. (35) of our first
paper in a simplified notation. Quantities with the \textit{zero} subscript
represent units or intervals\ (of space, time or mass) as measured by an
observer at his/her location and time (typically at the origin of space and at
time $t_{0}$ as usual). Quantities with the $z$ subscript represent the same
units or intervals as \textquotedblleft seen\textquotedblright\ by the
observer located at the origin, but when these \textquotedblleft
objects\textquotedblright\ are placed\ at a cosmological location
characterized by redshift $z$.\footnote{We have avoided so far this subscript
notation (also carefully avoided in paper I) because it is easily confused
with the standard cosmology notation, where the \textit{zero} subscript
normally indicates the observed (redshifted) quantity, as opposed to the
non-redshifted quantity (usually with no subscript). On the contrary, in our
new notation the \textquotedblleft redshifted\textquotedblright\ quantity $q$
acquires a $z$ subscript ($q_{z}\equiv q(t)$, observed at the origin, but with
information coming from past time $t$), while the \textquotedblleft
non-redshifted\textquotedblright\ quantity acquires the \textit{zero}
subscript ($q_{0}\equiv q(t_{0})$, observed at the origin, at current time
$t_{0}$). Since this subscript notation is much more compact than our previous
one, we will adopt it for the rest of this paper. For instance, we will write
the current values of our cosmological parameters as $\delta_{0}\equiv
\delta(t_{0})$, $\gamma_{0}\equiv\gamma(t_{0})$, etc.} Space-time intervals
are dilated\ by the $(1+z)$ factor, simply because this is our new
interpretation of the cosmological redshift.

As for the masses, we have left the dependence on $z$ totally undetermined,
assuming that some function $f(1+z)$ will connect units of mass at different
locations in the Universe. We only suppose that $f$ will be a function of the
redshift factor $(1+z)$, such that for $z=0$ it reduces to unity, i.e.,
$f(1)=1$, leaving masses unchanged\ at locations where $z=0$. We will
determine this function $f$ in the following.

The scaling\ of the fundamental mechanical quantities in Eq. (\ref{eqn2.19})
will be reflected in similar properties of any other mechanical quantity or
unit. We consider for example the case of energy, as the luminosity discussed
above is just energy emitted per unit time. Since $1\ erg=1\ g\ cm^{2}%
\ s^{-2}$, an amount of energy can be written as $\delta E\sim\delta m\ \delta
l^{2}\ \delta t^{-2}$ and it is immediate to check that energies will scale
like masses, following the last equation. We can therefore write:%

\begin{equation}
\delta E_{z}=f(1+z)\ \delta E_{0}, \label{eqn2.20}%
\end{equation}
which implies that the total energy emitted by a standard candle placed at
redshift $z$ would be measured by the observer at the origin to be different
from the total energy emitted by the same candle when placed at the origin.
Since these energies are emitted during some (finite) intervals of time we can
write $\delta E_{z}=%
{\textstyle\int}
L_{z}(t_{z})dt_{z}=f(1+z)\ \delta E_{0}=f(1+z)\int L_{0}(t_{0})dt_{0}$, where
the luminosities are also labeled like all the other quantities and refer to
times connected by the same dilation factor\ $t_{z}=t_{0}(1+z)$.\footnote{In
the standard interpretation $t_{0}$ would be considered as the time $t_{rest}$
in the candle's rest-frame of reference and $t_{z}$ as the time $t_{obs}$
measured by the observer who sees the candle \textquotedblleft
moving\textquotedblright\ due to the expansion of the Universe. Standard
relativistic time dilation would yield $t_{obs}=t_{rest}(1+z)$. This time
dilation effect, which has been observed in the evolution of type Ia SNe
(\cite{1996ApJ...466L..21L}, \cite{Riess:1997zu}, \cite{Foley:2005qu},
\cite{Blondin:2008mz}), is also present in our theory, although its
interpretation is the one given by our fundamental Eq. (\ref{eqn2.19}).} Since
the infinitesimal time intervals are similarly related, $dt_{z}=dt_{0}(1+z)$,
the two luminosities are connected in the following way:%

\begin{equation}
L_{z}(t_{z})=\frac{f(1+z)}{(1+z)}\ L_{0}(t_{0}). \label{eqn2.21}%
\end{equation}

Our new definition of \textit{luminosity distance} will therefore generalize
the original definition in Eq. (\ref{eqn2.16}), by using the actual luminosity
$L_{z}$, instead of $L_{0}$, to correct for the intrinsic changes in the
candle's energy output. To obtain the full expression of this distance we use
Eqs. (\ref{eqn2.18}) and (\ref{eqn2.21}) together:%

\begin{equation}
d_{L}=\sqrt{\frac{L_{z}}{4\pi l}}=\sqrt{\frac{f(1+z)}{(1+z)}\frac{L_{0}}{4\pi
l}}=\mathbf{R}_{0}\mathbf{r=R}_{0}\frac{\delta_{0}(1+z)+\sqrt{(1+z)^{2}%
-(1-\delta_{0}^{2})}}{(1-\delta_{0}^{2})}, \label{eqn2.22}%
\end{equation}
where we used our simplified notation for the parameters $\delta_{0}%
\equiv\delta(t_{0})$, $\mathbf{R}_{0}\equiv\mathbf{R}(\mathbf{t}_{0})$, as
previously discussed.

Comparing our new expression with the original one in Eq. (\ref{eqn2.16}), we
see that basically we replaced the $(1+z)$ factor with a more complex factor
$\sqrt{(1+z)/f(1+z)}$, where again the function $f(1+z)$ is still undetermined
at this point. This is because the standard\ definition would use the
\textquotedblleft constant\textquotedblright\ luminosity $L_{0}$, as
$d_{L}=\sqrt{\frac{L_{0}}{4\pi l}}=\sqrt{(1+z)/f(1+z)}\mathbf{R}_{0}%
\mathbf{r}$, instead of the \textquotedblleft variable\textquotedblright%
\ $L_{z}$. One could argue that if the $f$ function were to be equal to
$(1+z)^{-1}$, i.e., if masses were to scale like $(1+z)^{-1}$, we would
recover the original definition, but this is not exactly the case. Again, the
original definition (\ref{eqn2.16}) assumes an invariant value $L=L_{0}$ of
the standard candle's luminosity at all cosmological locations, while our new
definition (\ref{eqn2.22}) is based on the choice of $L_{z}$ as reference
standard candle's luminosity and in general this quantity is not invariant
anymore, but changes according to Eq. (\ref{eqn2.21}).

In other words, we could have defined instead:%

\begin{equation}
d_{L}=\sqrt{\frac{L_{0}}{4\pi l}}=\sqrt{(1+z)/f(1+z)}\mathbf{R}_{0}%
\mathbf{r}=\mathbf{\sqrt{(1+z)/f(1+z)}R}_{0}\frac{\delta_{0}(1+z)+\sqrt
{(1+z)^{2}-(1-\delta_{0}^{2})}}{(1-\delta_{0}^{2})}, \label{eqn2.22.1}%
\end{equation}
using $L_{0}$ instead of $L_{z}$, and then supplementing this definition with
the information of Eq. (\ref{eqn2.21}). We will see that this will not affect
the subsequent discussion on type Ia Supernovae. We prefer our definition in
Eq. (\ref{eqn2.22}) as it corrects the distance estimates, due to the
variability of the candle's luminosity. For example, if a candle is
intrinsically dimmed when placed far away, its distance should be smaller than
the one estimated with the original definition, given the same value $l$ for
the observed apparent luminosity. Our new definition (\ref{eqn2.22}) precisely
incorporates such corrections and will imply a revision of the current
distance estimates based on luminosity measurements.

It will be useful in the following to consider also the spectral energy
distribution, $F(\lambda)\equiv\frac{dL}{d\lambda}$. Following Eq.
(\ref{eqn2.21}), we can write $\int F_{z}(\lambda_{z},t_{z})d\lambda_{z}%
=\frac{f(1+z)}{(1+z)}\int F_{0}(\lambda_{0},t_{0})d\lambda_{0}$, where the
meaning of the subscripts is the same as in the preceding equations. The
wavelength variables and related differentials are connected as usual,
$\lambda_{z}=\lambda_{0}(1+z)$ and $d\lambda_{z}=d\lambda_{0}(1+z)$, so that
the direct relation between $F_{z}$ and $F_{0}$ is%

\begin{equation}
F_{z}(\lambda_{z},t_{z})=\frac{f(1+z)}{(1+z)^{2}}\ F_{0}(\lambda_{0},t_{0}).
\label{eqn2.23}%
\end{equation}

After discussing at length the modifications to the luminosity distance, we
return for completeness to the other definitions of cosmological distances.
Two other distances are usually introduced, the \textit{angular diameter
distance} $d_{A}$, when an extended light source of true proper diameter $D$
is placed at an (angular) distance $d_{A}$ and observed having an angular
diameter $\vartheta$, and the \textit{proper-motion distance} $d_{M}$, when
proper motions in the direction transverse to the line of sight are
considered. Their standard definitions are \cite{Weinberg}:%

\begin{align}
d_{A}  &  \equiv\frac{D}{\vartheta}=\mathbf{R(t)r=}\frac{\mathbf{R}%
_{0}\mathbf{r}}{(1+z)}\label{eqn2.24}\\
d_{M}  &  \equiv\frac{V_{\bot}}{\Delta\vartheta/\Delta t_{obs}}=\mathbf{R}%
_{0}\mathbf{r,}\nonumber
\end{align}
where, in the second definition, $V_{\bot}$ is the true velocity of the source
in the direction perpendicular to the line of sight and $\Delta\vartheta$ is
the change in the (angular) position of the object during the time interval of
observation $\Delta t_{obs}$.

Both definitions need to be reconsidered in our new interpretation. In the
first definition the \textquotedblleft true\textquotedblright\ diameter of the
light source will depend on $z$ as for all other lengths, $D_{z}=(1+z)D_{0}$,
but using the RW metric the proper distance across the source is
$D_{z}=\mathbf{R(t)r\ }\vartheta$ for small angular diameters $\vartheta\ll1$,
which leads essentially to the same expression as in the standard treatment,
just with $D$ replaced by $D_{z}$. Therefore, our definition of $d_{A}$ is
similar to the standard one, but needs to be supplemented by the scaling
equation of the source diameter:%

\begin{align}
d_{A}  &  \equiv\frac{D_{z}}{\vartheta}=\mathbf{R(t)r=}\frac{\mathbf{R}%
_{0}\mathbf{r}}{(1+z)}\label{eqn2.25}\\
D_{z}  &  =(1+z)D_{0}.\nonumber
\end{align}
In this way the right-hand side of the equation connecting $d_{A}$ to the
cosmological quantities is unaffected, but when we make contact with
observations, i.e., we use the left-hand side of the equation, the change of
the true diameter with $z$ must be included. Since $D_{z}$ increases with
(positive) $z$, using the old definition which assumes a fixed $D=D_{0}$,
would result in underestimating the diameter distances by a factor $(1+z)$.

Similar care has to be taken in revising the proper motion distance $d_{M}$. A
moving source will have\ a transverse velocity $V_{\bot}$ which is unaffected
by our Eq. (\ref{eqn2.19}), since it's a ratio between quantities which scale
in the same way. In the standard theory the time interval of observation
$\Delta t_{obs}$ is thought to be different from the time interval of motion
$\Delta t_{mot}$ due to the usual redshift factor, $\Delta t_{mot}=\Delta
t_{obs}/(1+z)$. However, in our view, the observed time $\Delta t_{obs}$ is
precisely the true time interval during which the object moved at a redshift
$z$: $\Delta t_{obs}=\Delta t_{z}$. The proper distance travelled is $\Delta
D_{z}=V_{\bot}\Delta t_{z}=V_{\bot}\Delta t_{obs}=\mathbf{R(t)r\ }%
\Delta\vartheta$, using the RW metric as in the treatment of the angular
distance above, and our revised expression of the proper motion distance is:%

\begin{equation}
d_{M}\equiv\frac{V_{\bot}}{\Delta\vartheta/\Delta t_{obs}}=\mathbf{R(t)r=}%
\frac{\mathbf{R}_{0}\mathbf{r}}{(1+z)}. \label{eqn2.26}%
\end{equation}

In this case, we had to modify the right-hand side of the equation (compared
to the standard definition), but the left-hand side is unaffected (in
particular $\Delta t_{obs}=\Delta t_{z}$ is still simply the observation time
interval). Finally, we note that our corrected expressions for $d_{A}$ and
$d_{M}$ are essentially the same (both lead to $\mathbf{R(t)r}$) since the
geometry is totally equivalent ($D_{z}$ is equivalent to $V_{\bot}\Delta
t_{obs}$, $\vartheta$ to $\Delta\vartheta$).

Equations (\ref{eqn2.14}), (\ref{eqn2.15}), (\ref{eqn2.22}), (\ref{eqn2.25}),
and (\ref{eqn2.26}) are our revised expressions for the classic distances used
in cosmology. As for their standard counterparts, they are all approximately
equal to each other for $z\ll1$ and $\mathbf{r}\ll1$, i.e., $d_{C}\simeq
d_{prop}\mathbf{(t}_{0}\mathbf{)}\simeq d_{P}\simeq d_{L}\simeq d_{A}\simeq
d_{M}\simeq\mathbf{R}_{0}\mathbf{r}$. We also note that three of the six
definitions of distance were modified by our new kinematical conformal
cosmology, thus potentially affecting current astronomical distance estimates.

One final consideration is needed regarding the SSC space coordinate $r$. This
dimensionful coordinate can be used to measure distances in the Static
Standard Coordinates, which are in general different from the cosmological
distances described above. However, for small distances, or for $z\ll1$, this
coordinate $r$ should also reduce to $\mathbf{R}_{0}\mathbf{r}$ as for all the
other distances. We recall from paper I that the coordinate transformation
between $r$ and $\mathbf{r}$ is%

\begin{equation}
r=\frac{1}{\sqrt{\left\vert k\right\vert }}\frac{\mathbf{r}}{\sqrt
{1+\mathbf{r}^{2}}-\delta_{0}\mathbf{r}}\longrightarrow\frac{\mathbf{r}}%
{\sqrt{\left\vert k\right\vert }}\simeq\mathbf{R(t}_{0}\mathbf{)r}
\label{eqn2.27}%
\end{equation}
where the limit in the last expression is for $\mathbf{r\ll0}$ and $\delta
_{0}\simeq0$. This last equation implies that the current value of the scale
factor is simply related to the parameter $k$:%

\begin{equation}
\mathbf{R}_{0}=\mathbf{R(t}_{0}\mathbf{)\simeq}\frac{1}{\sqrt{\left\vert
k_{0}\right\vert }}, \label{eqn2.28}%
\end{equation}
where, from now on, we will also denote with a zero subscript ($k_{0}$) the
current value of the $k$ parameter. Eq. (\ref{eqn2.28}) is particularly
important to simplify the connection between the Hubble constants $\mathbf{H}$
and $H$, in view of Eqs. (\ref{eqn2.6})-(\ref{eqn2.12}). It is easy to check
that, using the previous equation, the general connection simplifies to%

\begin{equation}
\mathbf{H}(\mathbf{t})\simeq H(t)(1+z), \label{eqn2.29}%
\end{equation}
so that for $z=0$ the two Hubble constants basically coincide, i.e.,
$\mathbf{H}_{0}\simeq H_{0}$.

It is beyond the scope of this paper to attempt a full revision of the
\textquotedblleft cosmological distance ladder,\textquotedblright\ in view of
our changes in the distance definitions. While we leave this analysis to
future work on the subject, we will concentrate our efforts in the next
section on our revised definition of the luminosity distance applied to type
Ia Supernovae.

\section{\label{sect:cosmological_parameters}Cosmological Parameters}

The central part of our analysis will deal with the evaluation of the current
values of the cosmological parameters in our model: the dimensionless $\delta$
parameter, the $\gamma$ and $k$ parameters (or the original $\kappa$ quantity,
see paper I \cite{Varieschi:2008fc}) all measured with reference to the
current time $t_{0}$.

We have already introduced $\gamma_{0}=\gamma(t_{0})\cong1.94\times
10^{-28}\ cm^{-1}$ in Sect. \ref{sect:hubble_constant}, but we still have to
show how this value was obtained and compare it to the original evaluation of
$\gamma$\ done by Mannheim. We will proceed first to estimate $\gamma_{0}$ in
the next sub-section, later we will obtain $\delta_{0}=\delta(t_{0})$\ from
Supernovae data and finally all the other parameters will be derived from
these two quantities.

\subsection{\label{sect:pioneer}Cosmological blueshift and the Pioneer
anomaly}

In Sect. \ref{sect:kinematical_conformal_cosmology} we have summarized all the
fundamental expressions of our cosmology and outlined the reasons why we
consider the $\mathbf{k}=-1$ solution as a possible description of the
evolution of the Universe. Although this solution can explain the observed
cosmological redshift, it has an additional new feature. It requires the
existence of a blueshift region in the immediate vicinity of our current
space-time position in the Universe.

This could be a serious setback for our model, since we normally do not
observe blueshift of nearby astrophysical objects except for the one caused by
the peculiar velocities of nearby galaxies, which is presumably due to
standard Doppler shift. However, experimental evidence has begun to accumulate
regarding a possible local region of blueshift, related to the so-called
\textit{Pioneer anomaly} (\cite{Anderson:1998jd}, \cite{Anderson:2001ks},
\cite{Anderson:2001sg}, \cite{Turyshev:2005zk}, \cite{Turyshev:2005ej},
\cite{Toth:2006qb}, \cite{Toth:2007uu}, \cite{Turyshev:2007ii},
\cite{Nieto:2005kb}, \cite{Nieto:2007tf}).

This is a small anomalous frequency drift, actually a blueshift, which was
observed analyzing the navigational data from the Pioneer 10/11 spacecraft,
received from distances between $20-70$ astronomical units from the Sun, while
exploring the outer Solar System. This \textquotedblleft
blueshift\textquotedblright\ frequency is uniformly changing at a rate of
$\overset{\cdot}{\nu}_{P}\cong6\times10^{-9}\ Hz/s$ and can be interpreted as
a constant sunward acceleration, reported as $a_{P}=-(8.74\pm1.33)\times
10^{-8}\ cm/s^{2}$ (radial inward acceleration) or as a \textquotedblleft
clock acceleration\textquotedblright\ $a_{t}\equiv\frac{a_{P}}{c}%
=-(2.92\pm0.44)\times10^{-18}\ s/s^{2}$, resulting in a frequency drift of
about $1.5\ Hz$ every $8$ years \cite{Anderson:2001ks}.

Preliminary findings indicate the possibility of detecting such anomaly also
in the radiometric data from other spacecraft traveling at the outskirts of
the Solar System, such as the Galileo and Ulysses missions
\cite{Anderson:2001ks}. This has prompted an extended re-analysis of all the
historical navigational data from these space missions, which is currently
underway (\cite{Turyshev:2005ej}, \cite{Toth:2006qb}, \cite{Toth:2007uu},
\cite{Turyshev:2007ii}), in order to determine additional characteristics of
the anomaly, such as its precise direction, the possible temporal evolution,
its dependence on heliocentric distance, etc. Future dedicated missions are
also being planned (\cite{Anderson:2002yc}, \cite{Nieto:2004np},
\cite{Dittus:2005re}) to test directly this puzzling phenomenon.

Meanwhile, the origin and nature of this anomaly remains unexplained; all
possible sources of systematic errors have been considered
(\cite{Anderson:2001ks}, \cite{Anderson:2001sg}, \cite{Turyshev:2005zk},
\cite{Toth:2007uu}, \cite{Turyshev:2007ii}, \cite{Nieto:2003rq}) but they
cannot account for the observed effect. Possible physical origins of the
signal were studied, including dark matter, modified gravity or other known
non-conventional theories, but no satisfactory explanation has been found so
far (see details in \cite{Anderson:2001ks}, \cite{Nieto:2005kb},
\cite{Nieto:2007tf}). On the contrary, we can analyze the Pioneer anomaly with
the model outlined in the previous sections and use the data reported above to
estimate the local values of our cosmological parameters.

The phenomenology of the Pioneer anomaly is related to the exchange of
radiometric data between the tracking station on Earth (Deep Space Network -
DSN) and the spacecraft, using S-band Doppler frequencies (in the range
$1.55-5.20\ GHz$). More precisely, an uplink signal is sent from the DSN to
the spacecraft at $2.11\ GHz$, based on a very stable hydrogen maser system,
the Pioneer then returns a downlink signal at a slightly different frequency
of about $2.29\ GHz$, to avoid interference with the uplink signal. This is
accomplished by an S-band transponder which applies an exact and fixed
turn-around ratio of $240/221$ to the uplink signal.

This procedure, known as a \textit{two-way Doppler coherent signal}, allows
for very precise tracking of the spacecraft since the returning signal is
compared to the original one, as opposed to a one-way Doppler signal (fixed
signal source on spacecraft, whose frequency cannot be directly monitored for
accuracy). This type of tracking system and the navigational capabilities of
the Pioneer spaceship (spin-stabilized spacecraft, power source of special
design, etc.) allowed for a great acceleration sensitivity of about
$10^{-8}\ cm/s^{2}$, once the influence of solar-radiation-pressure
acceleration decreased below comparable levels (for distances beyond about
$20\ AU$ from the Sun).

After a time delay of a few minutes or hours, depending on the distance
involved, the DSN station acquires the downlink signal and any difference from
its expected frequency is interpreted as a Doppler shift due to the actual
motion of the spacecraft. Modern-day deep space navigational software can also
predict with exceptional precision the expected frequency of the signal
returned from the Pioneer, which should coincide with the one observed on
Earth. On the contrary, a discrepancy is found, corresponding to the values
indicated above, whose origin cannot be traced to any systematic effect due to
either the performance of the spacecraft or the theoretical modeling of its navigation.

Moreover, the signal analysis performed so far (\cite{Anderson:2001sg},
\cite{Toth:2006qb}) indicates an almost constant value of the anomalous
acceleration or frequency shift reported above (temporal and space variation
of $a_{P}$ within $10\%$), over a range of heliocentric distances
$\sim20-70\ AU$, and possibly even closer distances $\lesssim10\ AU$.

The Pioneer phenomenology corresponds exactly to the simplest experiment we
might conceive to check the validity of our model. In principle, it would be
sufficient to set up a lab experiment in which we emit some radiation of known
wavelength $\lambda(t_{0})$ at time $t_{0}$, keep this radiation from being
absorbed for a long enough time and then compare its wavelength $\lambda
(t_{0})$ with the radiation emitted by the same source at a later time
$\lambda(t_{1})$, with $t_{1}>t_{0}$. In our model we would expect the two
wavelengths to be different and, if we are already in a phase of universal
contraction as illustrated by the red-solid curve of Fig. 5 in our paper I, we
would have $\lambda(t_{0})<\lambda(t_{1})$. In terms of frequencies,
$\nu(t_{0})>\nu(t_{1})$, i.e., the radiation from time $t_{0}$ would appear to
be blueshifted, compared to the same radiation emitted by the same source at a
later time $t_{1}$. We will now proceed to interpret the Pioneer anomaly in a
similar way, but we will return in section \ref{sect:consequences}\ on the
feasibility of detecting wavelength variation in lab experiments.

It is useful to introduce a new time coordinate $\overline{t}$: let
$\overline{t}=0$ be the time at which the radiometric signal is sent from
Earth to the spacecraft, which is then received at time $\overline{t}$ and
immediately retransmitted down to the DSN, arriving back on Earth at time
$2\overline{t}$. In the standard theory the model for the signal coming back
to Earth is based on the relativistic Doppler effect. We will denote as
$\nu_{\operatorname{mod}}$ the frequency of the expected signal following this
model. Its ratio with the reference frequency $\nu_{ref}$ of the signal (the
uplink frequency of about $2.11\ GHz$, since we consider the two-way system)
is therefore given by the standard relativistic Doppler formula (see Eq. 2.2.2
in \cite{Weinberg}):%

\begin{equation}
\frac{\nu_{\operatorname{mod}}}{\nu_{ref}}=\frac{\sqrt{1-\frac{\mathbf{v}^{2}%
}{c^{2}}}}{1+\frac{v_{r}}{c}}\simeq1-\frac{v_{r}}{c}, \label{eqn3.1}%
\end{equation}
where $v_{r}$ is the source radial velocity and the approximation on the
right-hand side holds to first order in $v_{r}/c$.

For the case of the Pioneer spacecraft we use $v_{r}=2v_{\operatorname{mod}%
}(\overline{t})$, where $v_{\operatorname{mod}}(\overline{t})$ is the expected
velocity of the spacecraft, according to the theoretical navigation model, at
time $\overline{t}$, when the spaceship receives and immediately re-transmits
the signal. The factor of two arises from the double Doppler shift involved
(two-way system). With this radial velocity, Eq. (\ref{eqn3.1}) to first order
in $v_{r}/c$ becomes:%

\begin{equation}
\nu_{\operatorname{mod}}(\overline{t})\simeq\nu_{ref}\left[  1-\frac
{2v_{\operatorname{mod}}(\overline{t})}{c}\right]  \label{eqn3.2}%
\end{equation}
and this frequency should be observed with high precision, due to the reported
excellent navigational control of the spacecraft.

The anomaly comes from observing a different frequency $\nu_{obs}(\overline
{t})>\nu_{\operatorname{mod}}(\overline{t})$, involving an additional
unexplained blueshift. Over the range of the observed anomaly, the frequency
difference is reported as:%

\begin{align}
\Delta\nu &  =\nu_{obs}(\overline{t})-\nu_{\operatorname{mod}}(\overline
{t})\simeq\overset{\cdot}{\nu}_{P}2\overline{t}\label{eqn3.3}\\
\overset{\cdot}{\nu}_{P}  &  \cong6\times10^{-9}\ Hz/s.\nonumber
\end{align}
We point out here that the cited references adopt a rather confusing
\textquotedblleft DSN sign convention\textquotedblright\ for the frequency
difference in Eq. (\ref{eqn3.3}) (see \cite{Toth:2006qb},
\cite{Anderson:2001sg} and Ref. (38) of \cite{Anderson:1998jd}), resulting in
a sign change in most of their equations. We prefer to use here our definition
of $\Delta\nu$ as given in the previous equation.

An alternative way of reporting the anomaly is the following. As in Eq.
(\ref{eqn3.2}) we can also write the observed frequency to first order in
$v_{r}/c$ as:%

\begin{equation}
\nu_{obs}(\overline{t})\simeq\nu_{ref}\left[  1-\frac{2v_{obs}(\overline{t}%
)}{c}\right]  \label{eqn3.4}%
\end{equation}
where the \textquotedblleft observed\textquotedblright\ velocity of the
spacecraft always refers to the time of interest $\overline{t}$. Combining
together these last three equations we write the frequency difference as:%

\begin{equation}
\Delta\nu=-2\frac{\nu_{ref}}{c}\left[  v_{obs}(\overline{t}%
)-v_{\operatorname{mod}}(\overline{t})\right]  . \label{eqn3.5}%
\end{equation}
We then multiply and divide the last equation by $\overline{t}$, so that we
can introduce $a_{P}=\left[  v_{obs}(\overline{t})-v_{\operatorname{mod}%
}(\overline{t})\right]  /\overline{t}$, the residual Pioneer acceleration of
unknown origin. This is the change of velocity over time $\overline{t}$ and
not over the double time $2\overline{t}$ as in Eq. (\ref{eqn3.3}). With this
residual acceleration, the Pioneer anomaly is usually reported as:%

\begin{align}
\Delta\nu &  =-2\frac{\nu_{ref}}{c}a_{P}\ \overline{t}\label{eqn3.6}\\
a_{P}  &  =-(8.74\pm1.33)\times10^{-8}\ \frac{cm}{s^{2}}\nonumber
\end{align}
where the negative sign of the residual acceleration indicates its sunward
direction, therefore we have a positive frequency shift corresponding to a
local unexplained blueshift of radiation emitted by the spacecraft.

Given our discussion in the preceding sections we can now explain Eqs.
(\ref{eqn3.6}) or (\ref{eqn3.3}) with our new interpretation. The Pioneer data
represent the equivalent of a local measurement of the current blueshift
predicted in Sect. \ref{sect:summary}, therefore they can be used to find the
value of the parameter $\gamma$ at the time when the Pioneer radiation was
emitted (a few years ago) but this can be considered to be the current value,
due to the slow variation of the cosmological parameters.

Let's consider the current time $t_{0}$ as the time of arrival of the Pioneer
radiation on Earth. The uplink signal was therefore sent at $t=t_{0}%
-2\overline{t}$ and retransmitted by the spacecraft, as a downlink signal, at
$t=t_{0}-\overline{t}$. We need to re-interpret Eqs. (\ref{eqn3.2}) and
(\ref{eqn3.4}) as, in our view, the anomaly is not due to a change in velocity
of the spacecraft (we assume $v_{\operatorname{mod}}(\overline{t}%
)=v_{obs}(\overline{t})=v(\overline{t})$), but just related to a shift of the
reference frequency. The reference frequency in Eq. (\ref{eqn3.4}) is the one
emitted in the past $\nu_{ref}=\nu(t=t_{0}-2\overline{t})$, while the
reference frequency in Eq. (\ref{eqn3.2}) is the one at current time
$\nu_{ref}=\nu(t=t_{0})$, so these two equations are modified respectively as follows:%

\begin{align}
\nu_{obs}(\overline{t})  &  \simeq\nu(t=t_{0}-2\overline{t})\ \left[
1-\frac{2v(\overline{t})}{c}\right] \label{eqn3.7}\\
\nu_{\operatorname{mod}}(\overline{t})  &  \simeq\nu(t_{0})\ \left[
1-\frac{2v(\overline{t})}{c}\right]  .\nonumber
\end{align}

Following Eq. (\ref{eqn3.3}) the frequency difference, to first order in
$v/c$, is now:%

\begin{equation}
\Delta\nu=\left[  \nu(t)-\nu(t_{0})\right]  \ \left[  1-\frac{2v(\overline
{t})}{c}\right]  =\nu(t)\left[  1-\frac{R(t_{0})}{R(t)}\right]  \ \left[
1-\frac{2v(\overline{t})}{c}\right]  \simeq\ \overset{\cdot}{\nu}_{P}%
(t_{0}-t), \label{eqn3.8}%
\end{equation}
having used our fundamental Eq. (35) from paper I and also $(t_{0}%
-t)=2\overline{t}$. We can simplify the equation above introducing additional
approximations. From the second line of Eq. (\ref{eqn2.2}) $1-\frac{R(t_{0}%
)}{R(t)}=1-\left\{  \cosh\left[  \sqrt{\left\vert k\right\vert }%
c(t_{0}-t)\right]  -\frac{\gamma}{2\sqrt{\left\vert k\right\vert }}%
\sinh\left[  \sqrt{\left\vert k\right\vert }c(t_{0}-t)\right]  \right\}
\cong\frac{\gamma}{2}c(t_{0}-t)$, since we can assume $\sqrt{\left\vert
k\right\vert }c(t_{0}-t)\ll1$.\footnote{The time delay for a two-way Pioneer
signal at the maximum distance of $70\ AU$ is about $20$ hours; we can assume
$\sqrt{\left\vert k\right\vert }\sim\gamma\sim10^{-28}-10^{-30}$, therefore
$\sqrt{\left\vert k\right\vert }c(t_{0}-t)\sim10^{-13}-10^{-15}\ll1$.} The
reference frequency is $\nu_{ref}=\nu(t)\cong\nu(t_{0})=2.11\ GHz$,
corresponding to the original uplink frequency \footnote{In some of the
references cited the reference frequency $\nu_{ref}$ is taken as the downlink
frequency of $2.29\ GHz$, since the downlink signal is compared directly with
this value. We prefer to use the uplink value, since this is the frequency
used at the original time $t=t_{0}-2\overline{t}$.} and we can also neglect
the ratio $2v(\overline{t})/c\ll1$, since the typical Pioneer speed in the
outer Solar System was $v\approx12\ km/s$ \cite{Anderson:2001sg}. With these
approximations the last equation becomes:%

\begin{equation}
\Delta\nu\cong\nu_{ref}\left[  \frac{\gamma}{2}c(t_{0}-t)\right]
\cong\ \overset{\cdot}{\nu}_{P}(t_{0}-t). \label{eqn3.9}%
\end{equation}

The clock acceleration $a_{t}$, the Pioneer acceleration $a_{P}$, the
frequency shift $\overset{\cdot}{\nu}_{P}$, and the reference frequency
$\nu_{ref}$ are all related together by:%

\begin{equation}
a_{t}=a_{P}/c=-\overset{\cdot}{\nu}_{P}/\nu_{ref} \label{eqn3.10}%
\end{equation}
which follows from Eq. (\ref{eqn3.3}) and (\ref{eqn3.6}). Combining these last
two equations and using the reported values of the Pioneer anomaly, we can
finally obtain our estimate of the current local value of the cosmological
parameter $\gamma$:%

\begin{equation}
\gamma_{0}=\gamma(t_{0})\cong\frac{2}{c}\frac{\overset{\cdot}{\nu}_{P}}%
{\nu_{ref}}=-\frac{2}{c^{2}}a_{P}=-\frac{2}{c}a_{t}=1.94\times10^{-28}%
\ cm^{-1}, \label{eqn3.11}%
\end{equation}
which represents the best estimate of $\gamma$ at our current time (although
the Pioneer data are a few years old). This is the value which was quoted in
Sect. \ref{sect:hubble_constant} and led to our evaluation of the Hubble
constant.\footnote{Since the discovery of the Pioneer anomaly many researchers
have noticed the numerical \textquotedblleft coincidence\textquotedblright%
\ between the Hubble constant and the value of the Pioneer acceleration
$a_{P}$ divided by $c$, and proposed many different explanations for this.
This coincidence is even more striking if one uses the value cited in Ref.
\cite{Anderson:2001sg} as the experimental value for Pioneer 10 data before
systematics, $a_{P}=-7.84\times10^{-8}\ cm\ s^{-2}$, thus obtaining
$H_{0}=80.7\ km\ s^{-1}\ Mpc^{-1}$ and $\gamma_{0}=1.74\times10^{-28}%
\ cm^{-1}$. Although this choice would result in a value of the Hubble
constant closer to standard cosmology evaluations, we prefer to base all our
analysis on the usually quoted value of $a_{P}$ as in Eq. (\ref{eqn3.6}).}

In the previous equation we connected $\gamma$ to the measured Pioneer
acceleration $a_{P}$ (or the clock acceleration $a_{t}$) simply because such
was the way these data were reported in the literature cited. However, it
should be clear from the discussion above that we explain the Pioneer anomaly
in terms of our \textit{cosmological-gravitational blueshift} (equivalent in a
way to the clock acceleration mentioned above). In this view, there is no real
dynamic acceleration of the Pioneer spacecraft (or of any other object in the
Solar System) oriented toward the Sun, due to some new gravitational force or
modification of existing gravity. In fact, $a_{P}=\left[  v_{obs}(\overline
{t})-v_{\operatorname{mod}}(\overline{t})\right]  /\overline{t}=0$ in our
analysis, since we assume there is no difference between the two velocities
$v_{obs}(\overline{t})$ and $v_{\operatorname{mod}}(\overline{t})$.

In this way we overcome the objection, reported for example in
\cite{Anderson:2001sg}, that \textquotedblleft the anomalous acceleration is
too large to have gone undetected in planetary orbits, particularly for Earth
and Mars,\textquotedblright\ since \textquotedblleft NASA's Viking mission
provided radio-ranging measurements \cite{1979ApJ...234L.219R}\ to an accuracy
of about $12\ m$,\textquotedblright\ which should have shown the effect of the
anomalous acceleration on the orbits of these two planets.

In fact, there cannot be any anomalous acceleration for the Earth or Mars, if
we measure distances with round-trip time of flight of radio signals
transmitted from Earth to the Viking spacecraft on the Mars surface
\cite{1977JGR....82.4329S}. On the contrary, we would observe a similar effect
for Earth, Mars, or any other object in the Solar System, if we were to study
its motion through Doppler frequency ranging, because of the intrinsic
differences in frequency or wavelength for light emitted at different
space-time positions, due to our new cosmological effects.

We will return on this difference between experimental techniques for
estimating distances in a later section. We simply conclude this part by
noting that if the Pioneer anomaly is indeed caused by our new cosmological
effects and not by modification of the gravity from the Sun, the
\textquotedblleft direction\textquotedblright\ of the anomalous
\textquotedblleft acceleration\textquotedblright\ should be pointing towards
the terrestrial observer and not towards the Sun. This is currently being
studied (see \cite{Toth:2006qb} for details) by using data from a period of
time when the spacecraft were much closer to the Earth and the Sun, so that a
clear direction of the anomaly can be determined, but no results from this new
analysis are available yet.

\subsection{\label{sect:supernovae}Kinematical Conformal Cosmology and type Ia
Supernovae}

The determination of $\delta_{0}$\ can be done with the powerful astrophysical
tool represented by type Ia Supernovae (SNe Ia) used as standard candles (see
\textquotedblleft The High-Z SN Search\textquotedblright\ web-site
\cite{HighZ}, or the \textquotedblleft High Redshift Supernova Search
Supernova Cosmology Project\textquotedblright\ web-page \cite{SCP}\ for an
introduction to the topic, see also \cite{Filippenko:1997ub},
\cite{Leibundgut:2001jd}, \cite{Perlmutter:2003kf}\ for reviews on the
subject). Since the original discoveries made by these two leading groups
(\cite{Perlmutter:1998np}, \cite{Riess:1998cb}) the observational data were
recently expanded to the so-called \textquotedblleft gold\textquotedblright%
\ and \textquotedblleft silver\textquotedblright\ sets\ (\cite{Riess:2004nr},
\cite{Riess:2006fw} and references therein) including the highest redshift
Supernovae known, at $z\gtrsim1.25$. From these data we will extract the value
of $\delta_{0}$ and then obtain all the other parameters.

The use of type Ia SNe as standard candles involves the revision of the
concept of luminosity distance and related quantities in view of our
alternative cosmology. The observational techniques must also be carefully
considered, since they are based on the current standard cosmology. We have
seen in the preceding section how to generalize the luminosity distance in
view of our new interpretation; we will now apply those concepts to the case
of type Ia SNe.

Eqs. (\ref{eqn2.19}) - (\ref{eqn2.23}) have shown that quantities such as the
absolute luminosity of standard candles and others, depend on their
cosmological location, simply described by the redshift parameter $z$. We want
to stress once more that in our new interpretation all the characteristic
quantities of emission and absorption of radiation, such as frequencies,
wavelengths, time intervals between radiative events, energies, etc., are
intrinsically dependent on the space-time location of the events being
observed. No receding motion is needed, no cosmological Doppler shift induces
redshifts in the observed spectra, but care is to be used to evaluate the
different quantities at the space-time location of interest and the proper
corrections will enter the formulas only when we compare the same quantity at
different cosmological locations.

For example, in our view the \textquotedblleft absolute\textquotedblright%
\ luminosity of a standard candle at position $\mathbf{r}$ emitting light at a
past time $\mathbf{t}$ would be measured differently by two observers, one
placed at the origin $\mathbf{r}=0$ observing the radiation at current time
$\mathbf{t}_{0}$ and the other at the source $\mathbf{r}$ observing the
radiation at time $\mathbf{t}$.\footnote{The absolute luminosity is the total
power emitted by the source. Obviously only apparent luminosities, i.e., power
per unit area, can be measured by observers far away from the source. The
absolute luminosity $L_{z}$ therefore refers to an ideal measurement, as if we
could measure all the energy flowing through a spherical surface of radius
$\mathbf{r}$, centered around the source.} Following our notation mentioned
above, the former observer would measure the luminosity $L_{z}$, which was
used in the previously mentioned equations to derive the luminosity distance.
The latter, observing the radiation at the source, would measure what we
denote by $L_{0}$, i.e., the equivalent of placing the standard candle near
the origin and observing it at our current time. The two luminosities $L_{z}$
and $L_{0}$\ are then related by Eq. (\ref{eqn2.21}).

To further clarify this issue with an example, a standard candle of
\textquotedblleft intrinsic\textquotedblright\ total luminosity $L_{0}=1%
\operatorname{erg}%
/%
\operatorname{s}%
$ and \textquotedblleft intrinsic\textquotedblright\ spectral color, e.g., a
\textquotedblleft blue\textquotedblright\ candle, will always have the same
luminosity and show its blue color when seen by an observer placed near the
candle, employing the appropriate local units of space, time, energy, etc. On
the contrary, when placed at some cosmological distance corresponding to some
definite $z$ value (say $z=1$ for example) it would be observed as a
\textquotedblleft red\textquotedblright\ candle and having a different
luminosity $L_{z}=\frac{f(1+z)L_{0}}{(1+z)}$ following Eq. (\ref{eqn2.21}),
using units proper to the observer at position $\mathbf{r}=0$ and time
$\mathbf{t}_{0}$. All these effects are just due to the different intrinsic
units used by the two observers, as related to Eq. (\ref{eqn2.19}) and are not
in any way connected to relativistic Doppler shifts, which might be only
additional corrections due to the peculiar motion of the source relative to
the observer.

The only missing piece in our luminosity definition is the explicit expression
of the function $f(1+z)$. We might expect this function to be related to the
usual $(1+z)$ factor as discussed before, but we have no reason to assume a
simple dependence such as the one followed by space-time intervals in Eq.
(\ref{eqn2.19}). In choosing the form for the function $f(1+z)$ we are guided
by the following considerations. The current theory of standard candles, as
already remarked before, considers a reference distance $d_{ref}=10\ pc$, at
which the apparent luminosity of the candle is taken as the absolute
luminosity. This leads to the standard expression of the luminosity distance
as in Eq. (\ref{eqn2.17}). Since our luminosity $L_{z}$ depends on $(1+z)$, it
seems more correct to place the candle at a position where $z=0$, in order to
estimate its \textquotedblleft absolute\textquotedblright\ luminosity. Instead
of having the candle at the origin of the space coordinates (a rather
impracticable choice if the candle is a Supernova) we can place it at a
distance $d_{rs}=\mathbf{R}_{0}\mathbf{r}_{rs}=\mathbf{R}_{0}\frac{2\delta
_{0}}{1-\delta_{0}^{2}}$, at the beginning of the \textquotedblleft redshift
region,\textquotedblright\ where it is also $z=0$ as discussed at length in
our first paper. In other words, there are only two positions where a standard
candle's luminosity is unaffected by the universal conformal stretching: one
is at the origin and the other, more conveniently used as a reference
distance, is this special distance $d_{rs}$, which also depends on the current
value $\delta_{0}$ of our fundamental parameter.

The second hypothesis that we will make is on the form of the function
$f(1+z)$. In this we are guided again by the classic definition of the
luminosity distance, based on the inverse square law. In fact, we propose a
generalization of the original inverse square law $l=\frac{L}{4\pi d^{2}}$
which assumed $L$ to be invariant, to a generalized form $l=\frac{L_{0}}{4\pi
d^{a}}$ where now $L_{0}$ is measured near the source, therefore constant by
definition, but we generalize the power dependence on the distance to account
for the luminosity variations with $z$. Since this form of the revised
inverse-square law\ would be dimensionally incorrect for $a\neq2$, we further
refine it by including the reference distance $d_{ref}=d_{rs}$ as follows:%

\begin{align}
l  &  =\frac{L_{z}(t_{z})}{4\pi d_{L}^{2}}=\frac{L_{0}(t_{0})}{4\pi d_{L}%
^{2}(d_{L}/d_{ref})^{a}}\label{eqn3.12}\\
L_{z}(t_{z})  &  =\frac{L_{0}(t_{0})}{(d_{L}/d_{ref})^{a}}\nonumber\\
f(1+z)  &  =(1+z)\frac{L_{z}(t_{z})}{L_{0}(t_{0})}=\frac{(1+z)}{(d_{L}%
/d_{ref})^{a}}=(1+z)\left[  \frac{2\delta_{0}}{\delta_{0}(1+z)+\sqrt
{(1+z)^{2}-(1-\delta_{0}^{2})}}\right]  ^{a}.\nonumber
\end{align}

In the equation above we show in the first line our \textquotedblleft
inverse-power law\textquotedblright\ generalization of the apparent
luminosity-distance relationship, with an exponent $a$ to be determined by
fitting the type Ia SNe data. This generalization implies the dependence on
$z$ of the absolute luminosity $L$, as given in the second line, and by
comparison with the Eq. (\ref{eqn2.21}) it determines the form of the unknown
function $f(1+z)$, as given in the last line of Eq. (\ref{eqn3.12}).

We want to emphasize here that our choice of the function $f(1+z)$ or of the
generalized \textquotedblleft inverse-power law\textquotedblright\ is just an
educated guess, which will lead to a good fit of the SNe data in the
following. At this stage we cannot justify it on the basis of our kinematical
conformal cosmology. Therefore, going back to our Eq. (\ref{eqn2.19}) which
details the scaling properties of units of length, time and mass, while the
first two lines are fundamentally based on our new interpretation, the last
expression with the function $f(1+z)$ of Eq. (\ref{eqn3.12}) is currently our
best hypothesis, but might need to be revised in the future.

In any case, assuming our current hypothesis to be a reasonable one, we can
finally obtain the equivalent of Eq. (\ref{eqn2.17}), which will express the
distance modulus $\mu(z)$ directly as a function of the redshift parameter.
Combining together Eqs. (\ref{eqn2.18}), (\ref{eqn2.22}) and (\ref{eqn3.12}),
we obtain:%

\begin{align}
\mu_{bol}(z;a,\delta_{0})  &  =m_{bol}(z,t_{z})-M_{bol}(z=0,t_{0}%
)=-2.5\log_{10}\left[  l_{bol}(z,t_{z})/l_{bol}(z=0,t_{0})\right]
=\label{eqn3.13}\\
&  =2.5(2+a)\log_{10}(d_{L}/d_{ref})=2.5(2+a)\log_{10}\left[  \frac{\delta
_{0}(1+z)+\sqrt{(1+z)^{2}-(1-\delta_{0}^{2})}}{2\delta_{0}}\right]  .\nonumber
\end{align}

In this equation we have indicated that all magnitudes, luminosities, etc.,
refer to bolometric quantities, i.e., are measured over all wavelengths of
emitted radiation. We will consider later the effect of observing this
radiation with particular filters. We have also indicated the time dependence
of the observed quantities, where as mentioned before $t_{z}=t_{0}(1+z)$. This
is due to the fact that type Ia SNe have a temporal evolution
(\cite{1996ApJ...466L..21L}, \cite{Riess:1997zu}, \cite{Foley:2005qu},
\cite{Blondin:2008mz}) and the luminosities are usually referred to the peak
values. The time at which a Supernova reaches its peak luminosity differs if
observed near the source ($t_{0}^{peak}$) or if it's seen from a cosmological
distance, in which case $t_{z}^{peak}=t_{0}^{peak}(1+z)$.

In Eq. (\ref{eqn3.13}) the (bolometric) distance modulus $\mu_{bol}%
(z;a,\delta_{0})$ is an explicit function of the redshift variable $z$ and of
the two parameters $a$ and $\delta_{0}$, whose values will be determined by
fitting this formula to the experimental data of type Ia SNe. Our treatment
therefore parallels the standard cosmology model where usually $\mu
_{bol}(z;\Omega_{M},\Omega_{\Lambda})$, as a function of the density
parameters $\Omega_{M}$, $\Omega_{\Lambda}$. The comparison with experimental
observations is usually carried out by fitting these expressions to the
measured distance moduli $\mu_{bol}$, for several type Ia SNe as observed by
the most recent surveys (\cite{Riess:2004nr}, \cite{Riess:2006fw}).

The determination of the bolometric magnitudes from the astrophysical
measurements is quite complex, involving conversions from observations
actually performed in precise wavelength bands ($U$, $B$, $V$, $R$, $I$
filters, for ultraviolet, blue, visible, red and infrared bands respectively)
to the total (bolometric) luminosities and corresponding magnitudes. This
involves precise \textquotedblleft K-corrections\textquotedblright\ to
transform the observations in the different bands to the bolometric
quantities, plus other corrections involving the extinction of the SNa light
in the host galaxy as well as in our own galaxy, resulting in a complex
procedure which is also able to discriminate the overall \textquotedblleft
quality\textquotedblright\ of the candidate Supernova. A more detailed
analysis of some of these procedures will be given in Appendix I, in order to
ascertain that they are consistent with our new cosmological interpretation
and with Eq. (\ref{eqn3.13}).

We have considered the best current compilation of existing data, given by the
292 SNe of the \textquotedblleft gold-silver set\textquotedblright\ taken from
Table 4 of Ref. \cite{Riess:2006fw} and also available in a machine-readable
form in Ref. \cite{riess2007}. In this table, as well as in similar data
compilations, the normalization of the distance moduli is usually arbitrary,
since only the relative distances are needed to obtain the dynamical
cosmological parameters, such as $\Omega_{M}$, $\Omega_{\Lambda}$. The overall
normalization of the data is linked to the value of the Hubble constant, but
this is usually treated as a nuisance parameter in the fitting procedure,
therefore not explicitly determined.

This approach follows from the standard cosmology definition of the luminosity
distance \cite{Carroll:1991mt} as a function of $z$, of the density parameters
$\Omega_{M}$, $\Omega_{\Lambda}$, and of the Hubble constant $H_{0}$:%

\begin{equation}
d_{L}(z;\Omega_{M},\Omega_{\Lambda},H_{0})=\frac{c(1+z)}{H_{0}\sqrt{\left\vert
\kappa\right\vert }}S\left\{  \sqrt{\left\vert \kappa\right\vert }%
\int\limits_{0}^{z}\left[  (1+z^{\prime})^{2}(1+\Omega_{M}z^{\prime
})-z^{\prime}(2+z^{\prime})\Omega_{\Lambda}\right]  ^{-1/2}dz^{\prime
}\right\}  , \label{eqn3.14}%
\end{equation}
where for $\Omega_{M}+\Omega_{\Lambda}>1$, $S(x)\equiv\sin(x)$ and
$\kappa=1-\Omega_{M}-\Omega_{\Lambda}$; for $\Omega_{M}+\Omega_{\Lambda}<1$,
$S(x)\equiv\sinh(x)$ and $\kappa$ as above; and for $\Omega_{M}+\Omega
_{\Lambda}=1$, $S(x)\equiv x$ and $\kappa=1$.

Since $d_{L}$ is inversely proportional to $H_{0}$, following Eq.
(\ref{eqn2.17}), the standard expression of the distance modulus becomes:%

\begin{align}
\mu(z)  &  =m(z)-M=5\log_{10}d_{L}(z;\Omega_{M},\Omega_{\Lambda}%
,H_{0})+25=\label{eqn3.15}\\
&  =5\log_{10}\mathcal{D}_{L}(z;\Omega_{M},\Omega_{\Lambda})-5\log_{10}\left(
H_{0}\right)  +25\nonumber
\end{align}
introducing the so-called \textquotedblleft Hubble
constant-free\textquotedblright\ luminosity distance $\mathcal{D}_{L}%
(z;\Omega_{M},\Omega_{\Lambda})\equiv H_{0}d_{L}$. The term containing the
Hubble constant is then usually summed together with $M$ and the other
constants in the previous equation and then integrated upon in the fitting procedure.

However, as noted in Ref. \cite{riess2007}, the data of the latest
\textquotedblleft gold-silver\textquotedblright\ sets can be reconciled with
the latest Cepheid-SNe based Hubble constant value \cite{Riess:2005zi}
($H_{0}=\left(  73\pm6.4\right)  \ km\ s^{-1}\ Mpc^{-1}$) by subtracting
$0.32\ mag$ from all distance moduli in Table 4 of \cite{riess2007}. By
performing this adjustment the distance moduli $\mu$ become consistent with
$H_{0}\cong73\ km\ s^{-1}\ Mpc^{-1}$, which is however different from our
value in Eq. (\ref{eqn2.9}), $H_{0}\cong89.7\ km\ s^{-1}\ Mpc^{-1}$.

Since standard cosmology assumes that the luminosity distance $d_{L}$ is
inversely proportional to $H_{0}$ as noted above, it is easy to check that
subtracting from all data another common value equal to $5\log_{10}%
(89.7/73)\cong0.45$, related to the ratio of the Hubble constants above, will
bring all the distance moduli in line with our preferred value of the Hubble constant.

Therefore, we have performed this \textquotedblleft
double-correction,\textquotedblright\ subtracting the factor $0.32+0.45=0.77$
from all the distance moduli of the \textquotedblleft
gold-silver\textquotedblright\ sets, to bring them in line with our
assumptions. This is the best we can do to fix the normalization of the
existing data in the literature, since we don't have access to the
normalization algorithm used by the SNe groups (see also the discussion in
Appendix I).

\begin{figure}
[ptb]
\begin{center}
\fbox{\ifcase\msipdfoutput
\includegraphics[
trim=0.000000in 0.000000in 0.002760in 0.000000in,
height=5.2607in,
width=6.301in
]%
{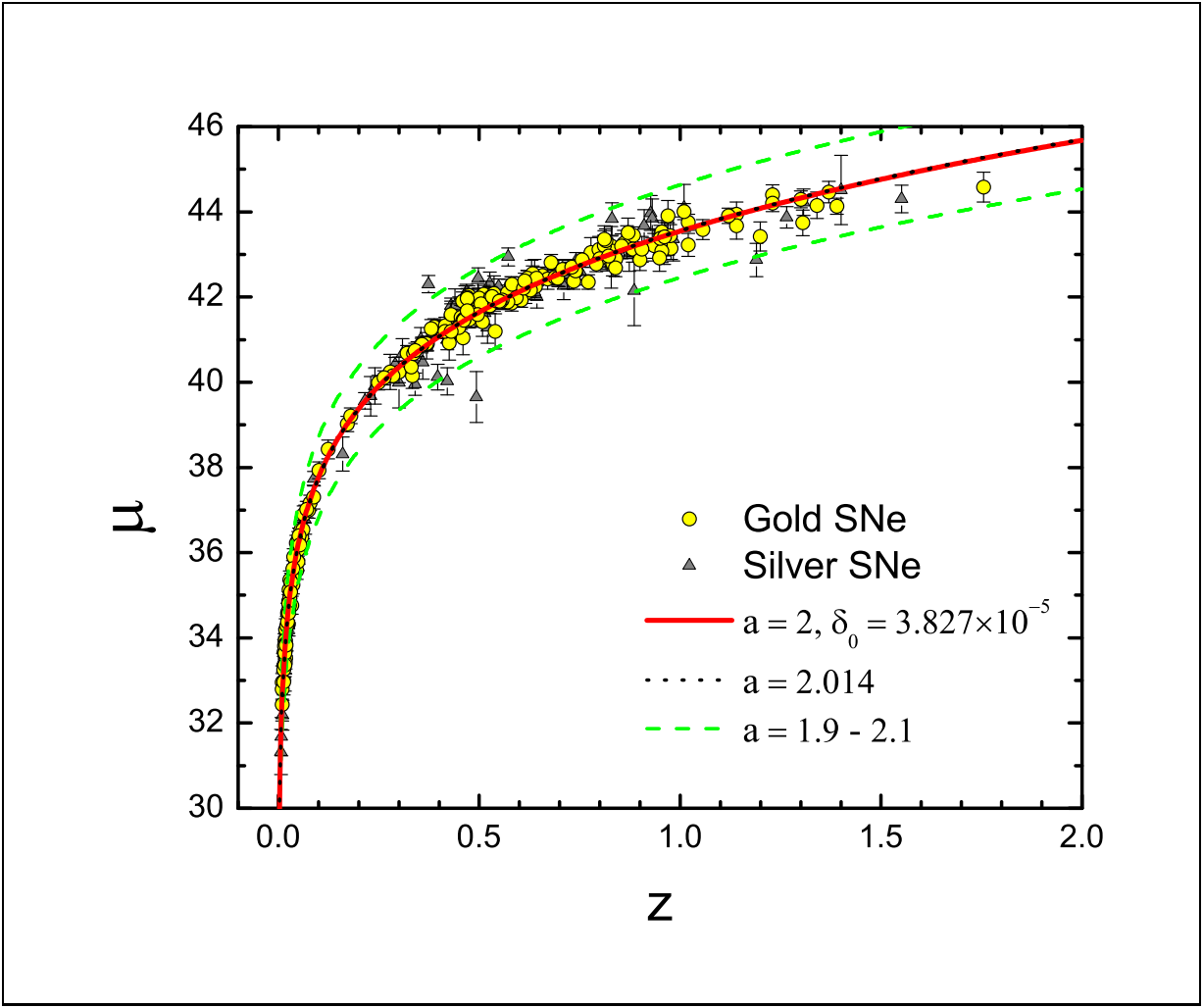}%
\else
\includegraphics[
height=5.2607in,
width=6.301in
]%
{C:/swp35/Docs/KINEMATICAL2/graphics/figure1__1.pdf}%
\fi
}\caption[Data from type Ia SNe \textquotedblleft gold-silver
sets\textquotedblright\ \cite{Riess:2006fw} are fitted with Eq. (\ref{eqn3.13}%
).]{Data from type Ia SNe \textquotedblleft gold-silver sets\textquotedblright%
\ \cite{Riess:2006fw} are fitted with Eq. (\ref{eqn3.13}). Our fits show a
remarkably good quality ($\chi^{2}=0.053$; $R^{2}=0.996$) for both a variable
$a$ (black dotted curve) and for a fixed $a=2$ (solid red curve). In this
latter case the best fit parameter for delta is $\delta_{0}=3.827\times
10^{-5}$. Also shown (dashed green curves) is the range of our fitting curves
for $a=1.9-2.1$.}%
\label{fig1}%
\end{center}
\end{figure}

We have plotted these \textquotedblleft double-corrected\textquotedblright%
\ distance moduli $\mu$ as a function of the observed redshift $z$ in Fig.
\ref{fig1} (yellow points for \textquotedblleft gold\textquotedblright\ data,
grey points for \textquotedblleft silver\textquotedblright\ SNe). In the same
figure we fit the expression in Eq. (\ref{eqn3.13}) to the experimental data,
choosing the \textquotedblleft gold set\textquotedblright\ only for our fit,
due to the greater reliability of these data over the \textquotedblleft silver
set.\textquotedblright\footnote{We have performed similar fits including both
\textquotedblleft gold\textquotedblright\ and \textquotedblleft
silver\textquotedblright\ data for completeness. We obtained basically the
same results as in the case of \textquotedblleft gold\textquotedblright\ only,
with a slightly worse statistical quality of the fit.} Our fit for the
\textquotedblleft gold\textquotedblright\ SNe has a good statistical quality
($\chi^{2}=0.0534$; $R^{2}=0.996$) and we obtain the best fit parameters as follows:%

\begin{align}
a  &  =2.014\pm0.018\label{eqn3.16}\\
\delta_{0}  &  =(3.951\pm0.167)\times10^{-5}.\nonumber
\end{align}

The resulting fit is shown as a black dotted curve in the figure. Given our
hypothesis of an \textquotedblleft inverse-power-law\textquotedblright\ our
fit seems to suggest an integer value $a=2$ for the exponent of the factor
$(d_{L}/d_{ref})$ in Eq. (\ref{eqn3.12}). If we perform a new fit of the same
data, keeping $a=2$ fixed, we obtain the continuous red curve in Fig.
\ref{fig1} (virtually equivalent to the black dashed one) and a slightly
different value for $\delta_{0}$:%

\begin{align}
a  &  =2\label{eqn3.17}\\
\delta_{0}  &  =(3.827\pm0.014)\times10^{-5}.\nonumber
\end{align}

Since this is also a good-quality fit of the data ($\chi^{2}=0.0533$;
$R^{2}=0.996$), we are inclined to consider the values given in Eq.
(\ref{eqn3.17}) as our best estimates of these parameters. In particular, we
will use in the following our inverse-power-law with an integer exponent
$a=2$. In Fig. \ref{fig1} we also plot the curves (green dashed) for
$a=1.9-2.1$, keeping $\delta_{0}$ as in Eq. (\ref{eqn3.17}), to show how our
fitting solution depends on the parameter $a$. It can be seen also that the
majority of the experimental points lies within this range, confirming our
hypothesis of $a\simeq2$.%

Our analysis of type Ia SNe therefore confirms the applicability of the
kinematical conformal cosmology to standard candle luminosity measurements and
proposes a small positive value $\delta_{0}\simeq3.83\times10^{-5}$ for our
fundamental cosmological parameter, as noted previously in this work and also
in paper I.

We have to mention again that the set of \textquotedblleft
gold-silver\textquotedblright\ SNe data and particularly the distance moduli
$\mu(z)$ from Ref. \cite{Riess:2006fw} that we used in our analysis, were
originally obtained through a rather complex calibration algorithm (called
MLCS2k2, see \cite{Riess:2004nr}, \cite{Riess:2006fw} for complete details)
which takes into consideration the wavelength bands being used ($U$, $B$, $V$,
$R$, $I$, etc.) and related K-corrections, the current value of the Hubble
constant, the extinction and reddening effects, the zero point luminosity
calibration and absolute magnitude of type Ia SNe, etc.

These calibration methods originated in the early papers of the two leading
groups, as the \textquotedblleft multicolor light-curve
shape\textquotedblright\ (MLCS - \cite{Riess:1994nx}, \cite{Riess:1996pa})
then revised into the latest MLCS2k2, and the template fitting method-$\Delta
m_{15}$ \cite{Phillips:1993ng}. In these early works it was still possible to
find a step-by-step description of the methods being used and the values of
almost all the corrections employed, in order to check our normalization procedure.

On the contrary, in the latest papers based on the MLCS2k2 method
(\cite{Riess:2005zi}, \cite{Riess:2004nr}, \cite{Riess:2006fw}), which
produced the \textquotedblleft gold-silver set\textquotedblright\ of distance
moduli that we used in Fig. \ref{fig1}, the complexity of the parametrization
and fitting procedure makes the comparison with our luminosity normalization
procedure very difficult. While we will analyze the K-corrections in Appendix
I, we make no attempt to revise the other corrections and normalization
procedures for the distance moduli (except for the \textquotedblleft
double-correction\textquotedblright\ we used above to bring the data in line
with our preferred value of the Hubble constant).

We are therefore aware that our fit of the \textquotedblleft
gold-silver\textquotedblright\ data based on Eq. (\ref{eqn3.13}) and presented
in the previous figure, might need some further adjustment to make it fully
consistent with the Supernova data, but this would not probably alter
significantly our previous discussion and the determination of our
cosmological parameters. We will leave a more detailed analysis and revision
of the calibration of Supernova data to future work.

\begin{figure}
[ptb]
\begin{center}
\fbox{\ifcase\msipdfoutput
\includegraphics[
height=5.2918in,
width=6.301in
]%
{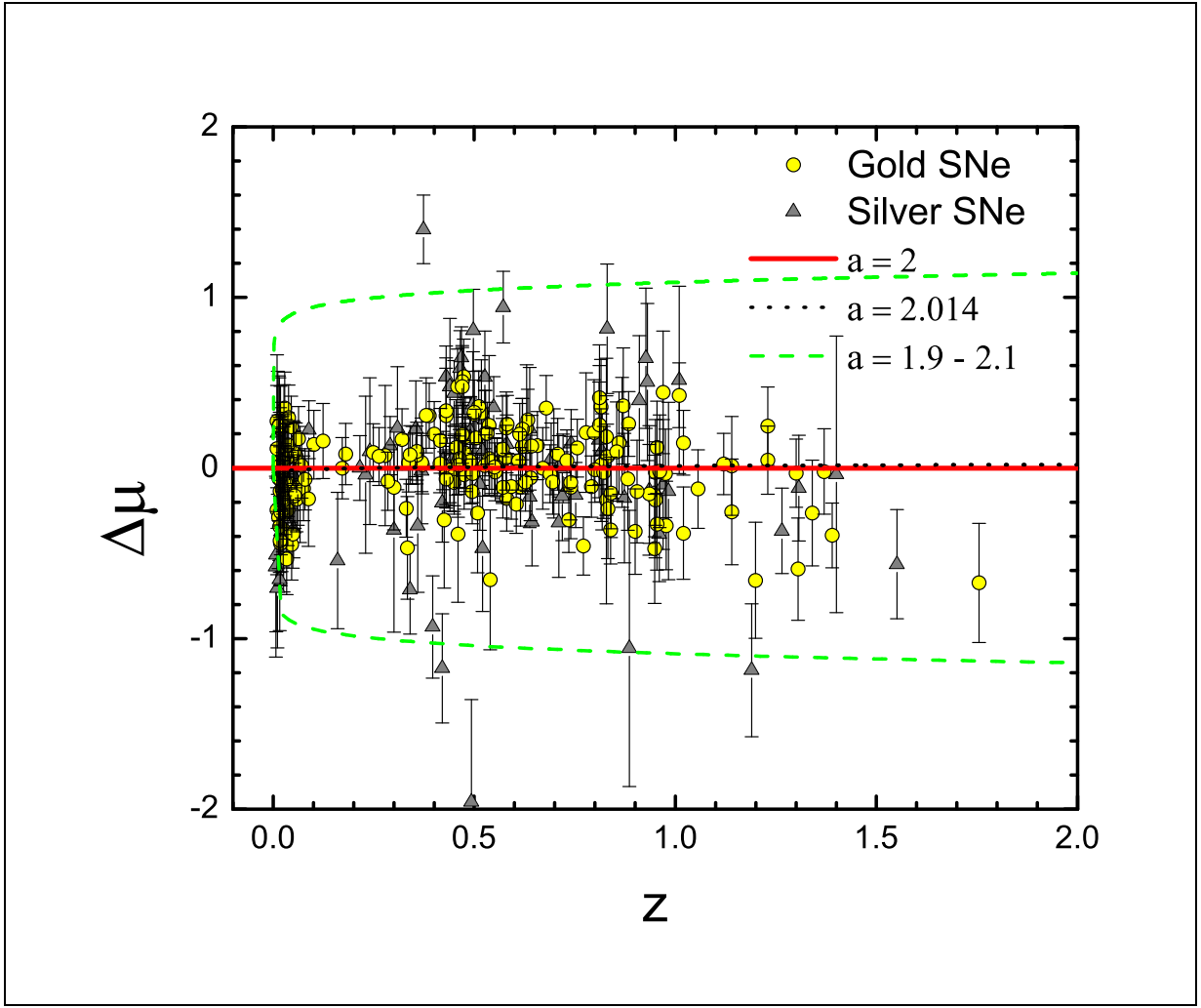}%
\else
\includegraphics[
height=5.2918in,
width=6.301in
]%
{C:/swp35/Docs/KINEMATICAL2/graphics/figure2__2.pdf}%
\fi
}\caption[Data from type Ia SNe \textquotedblleft gold-silver
sets\textquotedblright\ \cite{Riess:2006fw} are fitted with Eq. (\ref{eqn3.13}%
) and shown as residuals $\Delta\mu$.]{Data from type Ia SNe \textquotedblleft
gold-silver sets\textquotedblright\ \cite{Riess:2006fw} are fitted with Eq.
(\ref{eqn3.13}) and shown as residuals $\Delta\mu$. The baseline is
represented by our fit for fixed $a=2$ (solid red curve). The meaning of the
other curves and parameters is the same as in Fig. \ref{fig1}.}%
\label{fig2}%
\end{center}
\end{figure}

In figure \ref{fig2} we present the same data and fitting curves as in Fig.
\ref{fig1}, but in the form of residual values $\Delta\mu$, where the baseline
is represented by our fit (in solid-red) with the values of Eq. (\ref{eqn3.17}%
). It can be seen again how most of the SNe data fall within the $a=1.9-2.1$ band.%

Before proceeding to study the other cosmological parameters, we want to
analyze briefly the low-z case and the related fit to Supernova data, to
confirm the feasibility of our approach also at low redshift. In standard
cosmology the expression of the luminosity distance as a function of $z$, such
as the one described in Eq. (\ref{eqn3.14}), is not easily integrated so that
the low-z behavior is usually studied by expanding in Taylor series around
$z=0$. This procedure yields the well-known low-z expansion \cite{Weinberg}:%

\begin{equation}
d_{L}=\frac{c}{\mathbf{H}_{0}}\left[  z+\frac{1}{2}\left(  1-\mathbf{q}%
_{0}\right)  z^{2}+\mathcal{O}(z^{3})\right]  . \label{eqn3.18}%
\end{equation}

The first-order term corresponds to the original Hubble's law, $v_{r}\simeq
cz\simeq\mathbf{H}_{0}d_{L}$, where $v_{r}$ is the recessional velocity of the
galaxy, following Hubble's original explanation based on a pure Doppler
effect. The equivalent expansion for the distance modulus is:%

\begin{equation}
\mu(z)=25-5\log_{10}\mathbf{H}_{0}+5\log_{10}(cz)+1.086\left(  1-\mathbf{q}%
_{0}\right)  z+\mathcal{O}(z^{2}), \label{eqn3.19}%
\end{equation}
where $\mathbf{H}_{0}$ is measured as usual in $km\ s^{-1}\ Mpc^{-1}$ while
$c=299792.458\ km\ s^{-1}$. These expressions need to be corrected, due to our
new luminosity distance $d_{L}=\mathbf{R}_{0}\mathbf{r}$ from Eq.
(\ref{eqn2.22}), as opposed to the standard cosmology expression
$d_{L}=(1+z)\mathbf{R}_{0}\mathbf{r}$. We also need to include the reference
distance $d_{rs}=\mathbf{R}_{0}\mathbf{r}_{rs}=\mathbf{R}_{0}\frac{2\delta
_{0}}{1-\delta_{0}^{2}}$ at which we start observing the redshift.

Since our expression for $d_{L}$\ in Eq. (\ref{eqn2.22}) is an explicit
function of $z$, we can directly Taylor expand this function and obtain the
following result:\footnote{We note that if we were to correct the standard
expansion formula, as given in Ref. \cite{Weinberg} or \cite{Weinberg2}, we
would obtain: $d_{L}=\frac{c}{\mathbf{H}_{rs}}\left[  z-\frac{1}{2}\left(
1+\mathbf{q}_{rs}\right)  z^{2}+\mathcal{O}(z^{3})\right]  =\mathbf{R}%
_{0}\left[  \frac{1}{\left\vert \delta_{0}\right\vert }z-\frac{1}{2\left\vert
\delta_{0}\right\vert ^{3}}z^{2}+\mathcal{O}(z^{3})\right]  $. This is the
same result we would get by expanding our luminosity distance for a negative
value $\delta_{0}<0$. This is because standard cosmology does not include the
distance $d_{rs}$ in the derivation and this is equivalent to considering the
current value of $\delta$ as negative.}%

\begin{equation}
d_{L}=\mathbf{R}_{0}\left[  \frac{2\delta_{0}}{1-\delta_{0}^{2}}%
+\frac{1+\delta_{0}^{2}}{\delta_{0}\left(  1-\delta_{0}^{2}\right)  }%
z-\frac{1}{2\delta_{0}^{3}}z^{2}+\mathcal{O}(z^{3})\right]  , \label{eqn3.20}%
\end{equation}
from which we could derive an expansion for the distance modulus $\mu(z)$. The
problem with these type of expressions is that they have a very limited range
of validity in our revised cosmology, due to the small value of $\delta
_{0}\simeq3.83\times10^{-5}$ (or due to the large value of $\mathbf{q}%
_{0}=\frac{1}{\delta_{0}^{2}}-1\simeq6.82\times10^{8}$). It is easy to check
that, in our new expression for $d_{L}$ in the last equation, we could neglect
the $z^{2}$ and higher order terms only for $z\lesssim10^{-9}$, which is a
range of no practical interest. Therefore a low-z analysis similar to the one
of standard cosmology is actually not feasible in our new approach.

However, due to the low value of $\delta_{0}\simeq3.83\times10^{-5}$, we can
expand Eq. (\ref{eqn2.22}) in powers of $\delta_{0}$ around $\delta_{0}=0$ (or
just consider Eq. (\ref{eqn2.22}) for $\delta_{0}\rightarrow0$). With a Taylor
expansion we obtain:%

\begin{align}
d_{L}  &  =\mathbf{R}_{0}\left[  \sqrt{z^{2}+2z}+\left(  1+z\right)
\delta_{0}+\mathcal{O}(\delta_{0}^{2})\right] \label{eqn3.21}\\
\mu(z)  &  =2.5(2+a)\log_{10}\left\{  \frac{1-\delta_{0}^{2}}{2\delta_{0}%
}\left[  \sqrt{z^{2}+2z}+\left(  1+z\right)  \delta_{0}+\mathcal{O}(\delta
_{0}^{2})\right]  \right\}  .\nonumber
\end{align}
The leading terms at low redshift of the expressions in the previous equation
are the following (neglecting the $1-\delta_{0}^{2}$ factor in the second
line, due to the small value of $\delta_{0}$):%

\begin{align}
d_{L}  &  \simeq\mathbf{R}_{0}\sqrt{2z}\label{eqn3.22}\\
\mu(z)  &  \simeq2.5(2+a)\log_{10}\left(  \frac{\sqrt{2z}}{2\delta_{0}%
}\right)  .\nonumber
\end{align}

\begin{figure}
[ptb]
\begin{center}
\fbox{\ifcase\msipdfoutput
\includegraphics[
height=5.233in,
width=6.301in
]%
{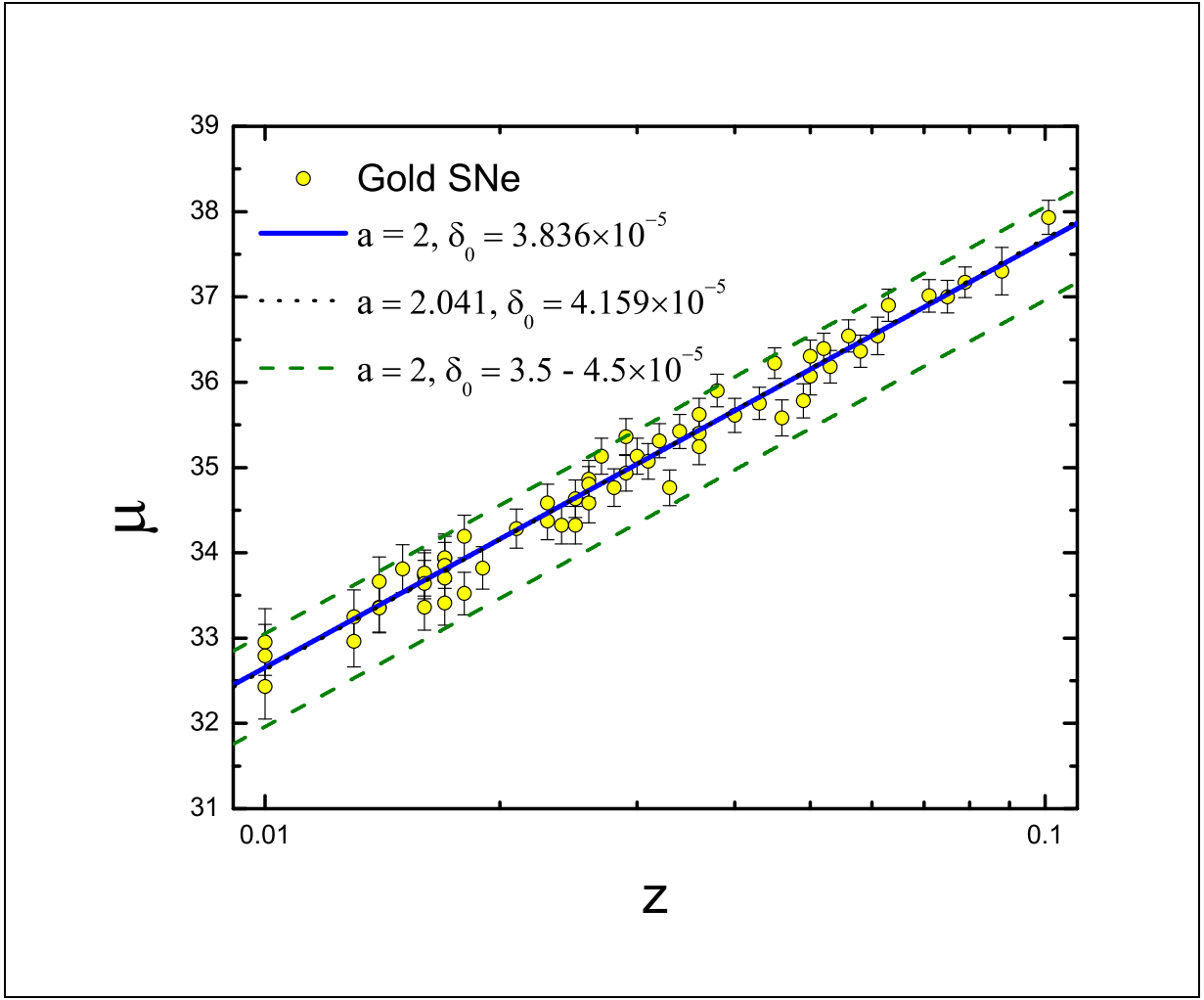}%
\else
\includegraphics[
height=5.233in,
width=6.301in
]%
{C:/swp35/Docs/KINEMATICAL2/graphics/figure3__3.pdf}%
\fi
}\caption[Data from type Ia SNe \textquotedblleft gold set\textquotedblright%
\ \cite{Riess:2006fw} are fitted with Eq. (\ref{eqn3.22}), our low-z
approximation.]{Data from type Ia SNe \textquotedblleft gold
set\textquotedblright\ \cite{Riess:2006fw} are fitted with Eq. (\ref{eqn3.22}%
), our low-z approximation. Our fits show a remarkably good quality ($\chi
^{2}=0.043$; $R^{2}=0.975$) for both a variable $a$ (black dotted curve) and
for fixed $a=2$ (solid blue curve). In this last case the best fit parameter
for delta is $\delta_{0}=3.836\times10^{-5}$. We also show (dashed green
curves) the range of our fitting expressions for $\delta_{0}=3.5-4.5\times
10^{-5}$. }%
\label{fig3}%
\end{center}
\end{figure}

This equation can be considered our \textquotedblleft low-z\textquotedblright%
\ expression and we can check its validity by using the second line to fit the
\textquotedblleft gold\textquotedblright\ SNe data of \cite{Riess:2006fw} in
the low-z regime. For this purpose we selected the \textquotedblleft
gold\textquotedblright\ data with $z\lesssim0.1$, with the same
\textquotedblleft double-correction procedure\textquotedblright\ mentioned
above and we used the last equation as a fitting formula. The results obtained
are the following:%

\begin{align}
a  &  =2.041\pm0.084\label{eqn3.23}\\
\delta_{0}  &  =(4.159\pm0.693)\times10^{-5},\nonumber
\end{align}
($\chi^{2}=0.0430$; $R^{2}=0.975$) leaving $a$ variable. If we constrain $a$
to our preferred value we obtain instead:%

\begin{align}
a  &  =2\label{eqn3.24}\\
\delta_{0}  &  =(3.836\pm0.023)\times10^{-5},\nonumber
\end{align}
which is also a good-quality fit of the data ($\chi^{2}=0.0425$; $R^{2}%
=0.975$). These results are shown in figure \ref{fig3}, where the fit with the
parameters of Eq. (\ref{eqn3.24}) is shown as a blue continuous curve, while
the black dotted line illustrates the fit with variable $a$. As in the
previous figures these two curves are virtually equivalent, confirming our
hypothesis of an integer value for $a$.%

In this figure we also show (dashed green curves) our low-z fitting curve of
Eq. (\ref{eqn3.22}) for $a=2$ and for the range $\delta_{0}=3.5-4.5\times
10^{-5}$, just to illustrate how sensitive our fitting formula is to the value
of $\delta_{0}$. Considering also the previous figures, we can see that the
value of the parameter $a$ determines the slope (or the shape of the curves in
a logarithmic plot), while the value of $\delta_{0}$ basically determines the
normalization of the curves, as it will also be shown in the following paragraphs.

Comparing these expressions with the standard cosmology ones in Eqs.
(\ref{eqn3.18}) - (\ref{eqn3.19}), it seems at first that our results do not
yield the standard Hubble's law, $v_{r}\simeq cz\simeq\mathbf{H}_{0}d_{L}$,
but as already remarked before this law was introduced by Hubble following the
original interpretation of the redshift as a pure (relativistic) Doppler
shift. This interpretation was later generalized into the cosmological
expansion, but this view is not shared by our model and therefore we don't
need to recover the original Hubble's law in our approach.

On the contrary, we can rewrite our \textquotedblleft low-z\textquotedblright%
\ expression for the distance moduli in Eq. (\ref{eqn3.22}), with $a=2$, as%

\begin{equation}
\mu(z)\simeq10\log_{10}\left(  \frac{\sqrt{2z}}{2\delta_{0}}\right)
=5\log_{10}\left(  \frac{z}{2\delta_{0}^{2}}\right)  \label{eqn3.25}%
\end{equation}
so that we have a perfect correspondence between our expression for $\mu(z)$
in the last equation and the classical one from Eq. (\ref{eqn3.19}), which can
also be rewritten as%

\begin{equation}
\mu(z)\simeq25+5\log_{10}\left[  \frac{cz}{\mathbf{H}_{0}}\right]  =5\log
_{10}\left[  10^{5}\frac{cz}{\mathbf{H}_{0}}\right]  , \label{eqn3.26}%
\end{equation}
neglecting higher order terms in $z$. Since both expressions (\ref{eqn3.25})
and (\ref{eqn3.26}) fit the experimental data and the standard one can be used
to measure the Hubble constant $\mathbf{H}_{0}$ (with $\mathbf{H}_{0}\simeq
H_{0}$, see discussion related to Eq. (\ref{eqn2.29})) comparing them together
and using the value of $\delta_{0}$ from Eq. (\ref{eqn3.24}) we obtain:%

\begin{equation}
\mathbf{H}_{0}\simeq H_{0}\simeq2\times10^{5}c\delta_{0}^{2}=88.2\ km\ s^{-1}%
\ Mpc^{-1}. \label{eqn3.27}%
\end{equation}

This value is very close to the Hubble constant we obtained in Sect.
\ref{sect:hubble_constant} (see Eq. (\ref{eqn2.9})) and that we used for our
calibration of the \textquotedblleft gold-silver\textquotedblright\ data. This
result shows that our calibration procedure of the SNe data was essentially
correct and links directly the Hubble constant to the fundamental parameter
$\delta_{0}$, as shown in the last equation.\footnote{We also tried fitting
the original low-z \textquotedblleft gold\textquotedblright\ data, based on
the standard value of $H_{0}=73\ km\ s^{-1}\ Mpc^{-1}$, proposed by Riess in
Ref. \cite{Riess:2005zi}. This resulted in a slightly different value of
$\delta_{0}\simeq3.461\times10^{-5}$, which is consistent with the Hubble
constant used, in view of Eq. (\ref{eqn3.27}).}

In addition, Eq. (\ref{eqn3.27}) can be combined with Eq. (\ref{eqn2.9}) to
give a direct relation between $\delta_{0}$ and $\gamma_{0}$:%

\begin{equation}
\gamma_{0}\simeq1.296\times10^{-19}\ \delta_{0}^{2}\ cm^{-1}, \label{eqn3.28}%
\end{equation}
where the numerical factor in the previous equation is a consequence of the
different units used to measure the Hubble constant. Using the value of
$\delta_{0}$\ from Eq. (\ref{eqn3.24}) or (\ref{eqn3.17}) into the last
equation, we obtain $\gamma_{0}\simeq1.9\times10^{-28}\ cm^{-1}$, as
introduced in Eq. (\ref{eqn3.11}) and based on the discussion of the Pioneer
anomaly.\footnote{We also performed a fit of the gold Supernova data using Eq.
(\ref{eqn3.21}), the \textquotedblleft low-$\delta$\textquotedblright%
\ expansion formula. Setting $a=2$, we obtained in this case $\delta
_{0}=3.868\times10^{-5}$, which placed in Eqs. (\ref{eqn3.27}) and
(\ref{eqn3.28}) yields respectively $H_{0}=89.7\ km\ s^{-1}\ Mpc^{-1}$ and
$\gamma_{0}=1.94\times10^{-28}\ cm^{-1}$, the exact values on which we base
our calibration of Supernova data. In this sense, our \textquotedblleft
low-$\delta$\textquotedblright\ expansion formula (\ref{eqn3.21}) should
probably take the place of the \textquotedblleft low-z\textquotedblright%
\ expression (\ref{eqn3.22}) as being more relevant to our analysis.}

\begin{figure}
[ptb]
\begin{center}
\fbox{\ifcase\msipdfoutput
\includegraphics[
height=5.2572in,
width=6.301in
]%
{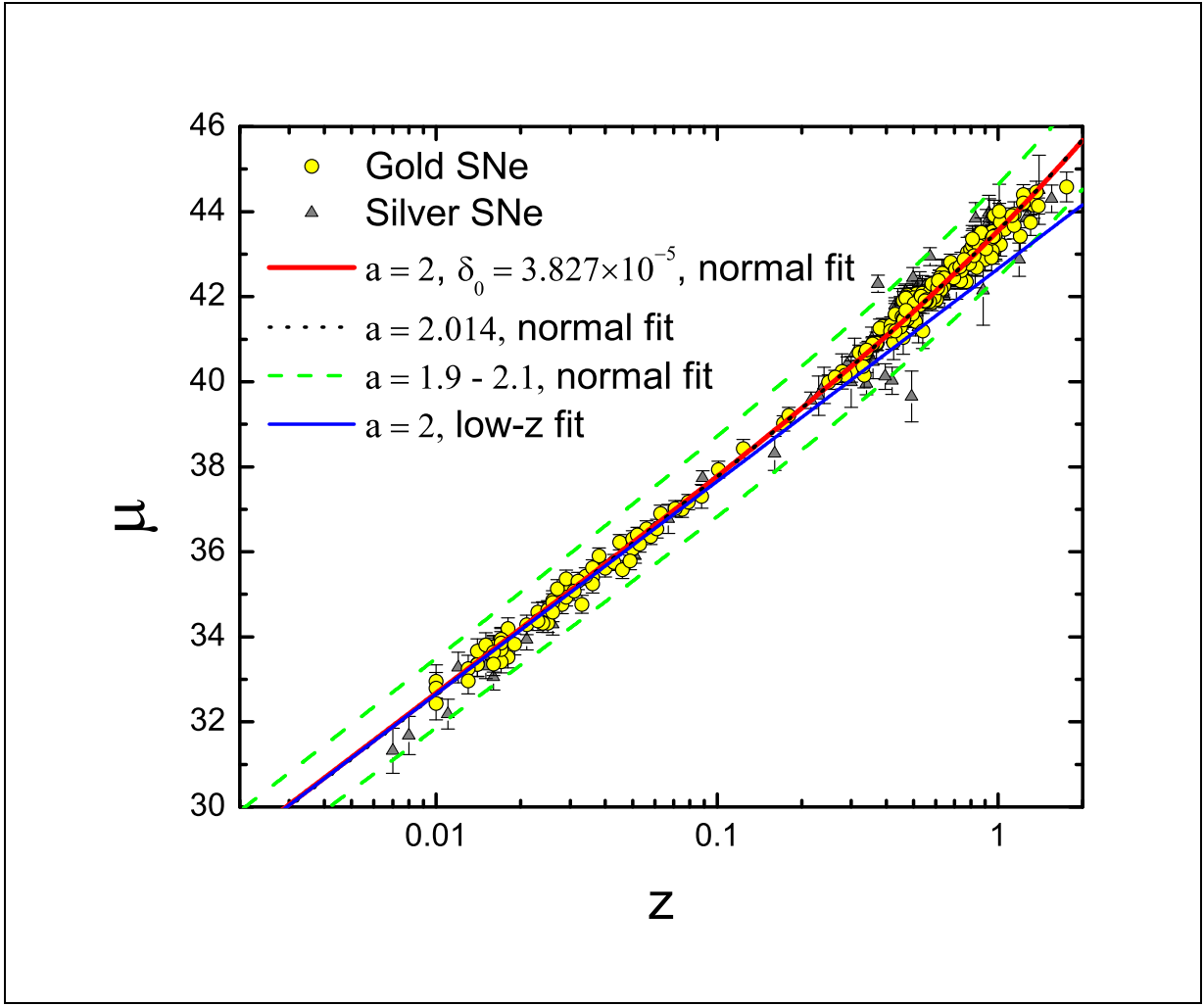}%
\else
\includegraphics[
height=5.2572in,
width=6.301in
]%
{C:/swp35/Docs/KINEMATICAL2/graphics/figure4__4.pdf}%
\fi
}\caption[Data from type Ia SNe \textquotedblleft gold-silver
sets\textquotedblright\ \cite{Riess:2006fw} are fitted with Eq. (\ref{eqn3.13}%
) - normal fit, and Eq. (\ref{eqn3.22}) - low-z fit.]{Data from type Ia SNe
\textquotedblleft gold-silver sets\textquotedblright\ \cite{Riess:2006fw} are
fitted with Eq. (\ref{eqn3.13}) - normal fit (red-solid curve for fixed $a=2$;
black dotted for variable $a$) and Eq. (\ref{eqn3.22}) - low-z fit (blue-solid
curve). Also shown by the green dashed curves is the range $a=1.9-2.1$ (normal
fit).}%
\label{fig4}%
\end{center}
\end{figure}

Finally, in figure \ref{fig4}, we reproduce the same data and fitting curves
as in the previous figures, but in the form of a standard linear Hubble\ plot,
to show that our approach can yield also these type of graphs, with the same
quality of those of standard cosmology.%

In particular, we see in this figure that our normal fit (red-solid curve)
following Eq. (\ref{eqn3.13}) and the \textquotedblleft
low-z\textquotedblright\ fit (in blue-solid) following Eq. (\ref{eqn3.22}) for
a fixed $a=2$ are almost equivalent at low redshift as expected, but they
become consistently different for $z\gtrsim0.1$. Since our normal fit, based
on $\delta_{0}=3.827\times10^{-5}$, can describe in a better way all the
Supernova data in our gold set, we will prefer this value over the low-z
evaluation of Eq. (\ref{eqn3.24}). We will therefore use $\delta_{0}%
\simeq3.83\times10^{-5}$ in the summary of the fundamental parameters
presented in the following section.

\subsection{\label{sect:other_parameters}The other cosmological parameters}

Using the results from the previous sections we are finally able to report our
best estimates of the parameters which enter our kinematical conformal
cosmology. These parameters are shown in Table \ref{TableKey:table1}, where
all the quantities either refer to their current value (subscript
\textit{zero}) or to the\ values at the location where the redshift starts
being observed (subscript $rs$) and we used for the estimates the value of
$\delta_{0}$ from Eq. (\ref{eqn3.17}).%

\begin{table}[tbp] \centering
\begin{tabular}
[c]{||l||}\hline\hline
\textbf{Kinematical Conformal Cosmology parameters}\\\hline
$\delta_{0}=3.83\times10^{-5}$\\\hline
$\gamma_{0}=1.94\times10^{-28}\
\operatorname{cm}%
^{-1}$\\\hline
$k_{0}=-\frac{\gamma_{0}^{2}}{4\delta_{0}^{2}}=-6.42\times10^{-48}\
\operatorname{cm}%
^{-2}$\\\hline
$\kappa_{0}=-\frac{\gamma_{0}^{2}}{4}-k_{0}=6.42\times10^{-48}\
\operatorname{cm}%
^{-2}$\\\hline
$\mathbf{H}_{0}\cong H_{0}=-\frac{\gamma_{0}}{2}c=-100\ h_{0}\ \frac
{km}{s\ Mpc}=-3.24\times10^{-18}\ h_{0}\ s^{-1}=-2.91\times10^{-18}\ s^{-1}%
$\\\hline
$h_{0}=0.897$\\\hline
$\mathbf{H}_{rs}\cong H_{rs}=+\frac{\gamma_{0}}{2}c=100\ h_{0}\ \frac
{km}{s\ Mpc}=3.24\times10^{-18}\ h_{0}\ s^{-1}=2.91\times10^{-18}\ s^{-1}%
$\\\hline
$q_{0}=q_{rs}=\frac{1}{\delta_{0}^{2}}-2=6.83\times10^{8}$\\\hline
$\mathbf{q}_{0}=\mathbf{q}_{rs}=\frac{1}{\delta_{0}^{2}}-1=6.83\times10^{8}%
$\\\hline
$\mathbf{R}_{0}=\mathbf{R(t}_{0})\simeq\frac{1}{\sqrt{\left\vert
k_{0}\right\vert }}=3.95\times10^{23}\ cm=0.128\ Mpc$\\\hline
$R_{0}=R\mathbf{(}t_{0})\simeq\mathbf{R(t}_{0}\mathbf{)}\sqrt{\left\vert
k_{0}\right\vert }\simeq1$\\\hline\hline
\end{tabular}
\caption{The fundamental parameters of our Kinematical Conformal Cosmology are shown here, as derived from the astrophysical data analyzed in the current section.}\label{TableKey:table1}%
\end{table}%

As it was previously mentioned, it is beyond the scope of this paper to
perform a full revision of the \textquotedblleft cosmological distance
ladder,\textquotedblright\ in view of the changes proposed by our new
approach. However, we want to show here the difference between our luminosity
distance estimates and the standard results, for different redshift values.
Table \ref{TableKey:table2} illustrates the results of this computation for
different values of the redshift parameter: in the second column we use our
new definition of luminosity distance following Eq. (\ref{eqn2.18}), while in
the third column we employ the standard cosmology formula in Eq.
(\ref{eqn3.14}) with $\Omega_{M}=0.27$, $\Omega_{\Lambda}=0.73$,
$H_{0}=89.7\ km\ s^{-1}\ Mpc^{-1}$ to bring it in line with our preferred
value of the Hubble constant.

As we can see from the table, there is a quite large difference between the
estimates in these two columns, a difference of about three orders of
magnitude just at low redshift, for $z\sim0.01-0.1$. This is mainly due to our
hypothesis of a change in the \textquotedblleft absolute\textquotedblright%
\ luminosity of standard candles in Eq. (\ref{eqn2.21}), resulting in
dramatically smaller revised distances.

However, if we were to redefine our new luminosity distance as in Eq.
(\ref{eqn2.22.1}), i.e., considering as in standard cosmology an invariable
absolute luminosity $L_{0}$ of the source and invoking corrections due to an
expansion of the Universe similar to standard cosmology we would obtain:%

\begin{equation}
d_{L}=\sqrt{\frac{L_{0}}{4\pi l}}=\sqrt{(1+z)/f(1+z)}\mathbf{R}_{0}%
\mathbf{r=R}_{0}\frac{\left[  \delta_{0}(1+z)+\sqrt{(1+z)^{2}-(1-\delta
_{0}^{2})}\right]  ^{2}}{2\delta_{0}(1-\delta_{0}^{2})}, \label{eqn3.30}%
\end{equation}
where we used our explicit form of the function $f(1+z)$ in Eq. (\ref{eqn3.12}%
) with $a=2$.%

\begin{table}[tbp] \centering
\begin{tabular}
[c]{||c|l|c|l|c||}\hline\hline
$z$ & $%
\begin{array}
[c]{c}%
d_{L}(Mpc)\\
\text{Eq.(\ref{eqn2.18})}\\
H_{0}=89.7\ \frac{km}{s\ Mpc}%
\end{array}
$ & $%
\begin{array}
[c]{c}%
d_{L}(Mpc)\\
\text{Eq.(\ref{eqn3.14})}\\
H_{0}=89.7\ \frac{km}{s\ Mpc}%
\end{array}
$ & $%
\begin{array}
[c]{c}%
d_{L}(Mpc)\\
\text{Eq.(\ref{eqn3.30})}\\
H_{0}=89.7\ \frac{km}{s\ Mpc}%
\end{array}
$ & $%
\begin{array}
[c]{c}%
d_{L}(Mpc)\\
\text{Eq.(\ref{eqn3.14})}\\
H_{0}=73\ \frac{km}{s\ Mpc}%
\end{array}
$\\\hline
$0.001$ & \multicolumn{1}{|c|}{$5.\,73\times10^{-3}$} & $3.35$ &
\multicolumn{1}{|c|}{$3.35$} & $4.11$\\\hline
$0.01$ & \multicolumn{1}{|c|}{$1.81\times10^{-2}$} & $33.7$ &
\multicolumn{1}{|c|}{$33.6$} & $41.4$\\\hline
$0.1$ & \multicolumn{1}{|c|}{$5.86\times10^{-2}$} & $360$ &
\multicolumn{1}{|c|}{$351$} & $442$\\\hline
$1$ & \multicolumn{1}{|c|}{$0.221$} & $5.25\times10^{3}$ &
\multicolumn{1}{|c|}{$5.02\times10^{3}$} & $6.45\times10^{3}$\\\hline
$10$ & \multicolumn{1}{|c|}{$1.40$} & $8.41\times10^{4}$ &
\multicolumn{1}{|c|}{$2.01\times10^{5}$} & $1.03\times10^{5}$\\\hline
$100$ & \multicolumn{1}{|c|}{$12.9$} & $1.04\times10^{6}$ &
\multicolumn{1}{|c|}{$1.71\times10^{7}$} & $1.27\times10^{6}$\\\hline
$1000$ & \multicolumn{1}{|c|}{$128$} & $1.11\times10^{7}$ &
\multicolumn{1}{|c|}{$1.68\times10^{9}$} & $1.37\times10^{7}$\\\hline\hline
\end{tabular}
\caption{Comparison between luminosity distances in our model and in standard cosmology, for different values of the redshift parameter.}\label{TableKey:table2}%
\end{table}%

In the fourth column of Table \ref{TableKey:table2}, we used the previous
equation to compute the distances and we notice that they are close to the
values of standard cosmology for both $H_{0}=89.7\ km\ s^{-1}\ Mpc^{-1}$
(third column) or for the more standard value $H_{0}=73\ km\ s^{-1}\ Mpc^{-1}$
(values in the last column of the table). This is of course expected, since
the distance in Eq. (\ref{eqn3.30}) would also fit the Supernova data, if we
were to assume an expansion equivalent to standard cosmology. In other words,
the last equation is our equivalent of the \textquotedblleft
standard\textquotedblright\ luminosity distance and yields to virtually the
same distance estimates as in standard cosmology. However, in our
interpretation Eq. (\ref{eqn2.18}) is to be considered the correct distance,
since it includes the intrinsic dimming of the source.

We wanted to introduce also this \textquotedblleft
standard-equivalent\textquotedblright\ luminosity distance in Eq.
(\ref{eqn3.30}), in order to make a comment on the so-called \textquotedblleft
Tolman surface brightness\ test,\textquotedblright\ which is usually employed
in cosmology to distinguish between standard expansion theories and
alternative models of redshift, such as the \textquotedblleft tired
light\textquotedblright\ explanation (see discussion in sect 1.7 of Ref.
\cite{Weinberg2} and recent experimental results in \cite{2001AJ....122.1084L}%
). This test is based on the ratio between the angular diameter distance
$d_{A}$ and the luminosity distance $d_{L}$. Using standard cosmology
distances from Eqs. (\ref{eqn2.16}) and (\ref{eqn2.24}) this ratio is simply
$d_{A}/d_{L}=(1+z)^{-2}$ so that the surface brightness $B$ of a luminous
object (defined as the apparent luminosity per solid angle - $l/\Omega$) will
result in $B\equiv\frac{l}{\Omega}=\frac{\mathcal{L}}{4\pi}\left(  \frac
{d_{A}}{d_{L}}\right)  ^{2}=\frac{\mathcal{L}}{4\pi}\frac{1}{\left(
1+z\right)  ^{4}}$, where $\mathcal{L}$ is the intrinsic absolute luminosity
per unit of proper area of the source (see \cite{Weinberg2}\ for details). As
shown in the above equation, this quantity should scale like $(1+z)^{-4}$ and
this prediction is recovered (within certain limits, due to the evolution of
the galactic light sources) in experimental studies \cite{2001AJ....122.1084L}%
. Tired light theories would require $B$ to scale as $(1+z)^{-1}$ and are
essentially ruled out by these experimental evidences.

In our model the ratio $d_{A}/d_{L}$, constructed using Eqs. (\ref{eqn2.18})
and (\ref{eqn2.25}), doesn't seem to scale as required by the Tolman
brightness test, i.e., $d_{A}/d_{L}=(1+z)^{-2}$, but rather as $(1+z)^{-1}$,
yielding a surface brightness scaling as $(1+z)^{-2}$. However, experimental
tests of this effect, such as the one reported in Ref.
\cite{2001AJ....122.1084L}, are based on a standard approach using invariable
luminosities and invariable diameters of the light sources being studied.
Following the discussion in the previous paragraphs, this amounts to using our
\textquotedblleft standard-equivalent\textquotedblright\ luminosity distance
in Eq. (\ref{eqn3.30}), instead of the one in Eq. (\ref{eqn2.18}). We can see
from the fourth column in our Table \ref{TableKey:table2} that distances
\ computed with this \textquotedblleft standard-equivalent\textquotedblright%
\ expression are similar to those calculated by standard cosmology. Therefore,
we infer that current tests of the Tolman effect would not be in disagreement
with our model. We will leave to future work a more detailed analysis of this effect.

In any case, if our approach is correct and we use our new luminosity distance
in Eq. (\ref{eqn2.18}), other distance estimates in the cosmological ladder
might also need to be revised. We mentioned in Sect.
\ref{sect:luminosity_distance}, that several other distances need to be
changed and that for example our new angular diameter distance would imply
larger distances than previously thought, therefore the overall reduction in
distance estimates might be less dramatic than the one illustrated in Table
\ref{TableKey:table2}, comparing just the second and third columns.

This is related to the final point we want to address in this section, i.e.,
the apparent discrepancy between our estimate of the parameter $\gamma
_{0}=1.94\times10^{-28}\
\operatorname{cm}%
^{-1}$ and the value proposed by Mannheim, $\gamma_{Mannheim}=3.06\times
10^{-30}%
\operatorname{cm}%
^{-1}$ \cite{Mannheim:1996rv}. We recall that this value was obtained by
Mannheim using a sample of eleven galaxies, where the rotational motion data
were fitted by the conformal gravity theory over a range of radial distances
of a few kiloparsec, from the center of each galaxy. The non-Keplerian effects
are manifest beyond the peak value at $r_{peak}=2.2\ r_{0}$, with $r_{0}$
ranging from $0.48\ kpc$ to $4.48\ kpc$, for the sample of eleven galaxies.
The average $r_{peak}$ is about $4.27\ kpc$, therefore the measured
$\gamma_{0}$ was obtained by fitting the original Mannheim-Kazanas potential
over radial distances $r\gtrsim r_{peak-ave}=4.27\ kpc$ from the reference
point of observation (the center of each galaxy). The global redshift of each
galaxy was already subtracted from the rotational data, therefore the measured
$\gamma_{0}$ refers to the intrinsic scale of the galaxies being considered
(kiloparsec scale) and should be in line with the value we propose.

However, Mannheim's analysis was based on standard cosmology estimates of the
distances to all these galaxies. We have seen above that our new
interpretation of the luminosity distance implies rather smaller distances
than those previously estimated. The ratio between the two estimates of the
gamma parameter%

\begin{equation}
\gamma_{0}/\gamma_{Mannheim}=63.4 \label{eqn3.31}%
\end{equation}
could be explained in terms of a similar ratio between distance estimates in
the standard theory vs. our new approach. This is due to the fact that, as
mentioned in our paper I, galactic rotational curve were fitted by Mannheim
using a potential proportional to the quantity $\gamma r$, so that the
overestimation of the distances $r$ would result in an underestimation of the
$\gamma$ parameter. We will also leave to future work a more detailed analysis
of this issue.

\section{\label{sect:consequences}Consequences of the Model}

To conclude the analysis of our kinematical conformal cosmology, we want to
summarize in this section the scaling properties of all the dimensionful
quantities, which could be linked to the more general problem of the time
variation of the physical constants. In our view, even more important is the
analysis of the dimensionless parameters and constants, which leads us to
establish a direct connection between our parameter $\delta$ and the
fine-structure constant $\alpha_{em}$ of the electromagnetic theory.

\subsection{\label{sect:scaling}Scaling properties of dimensionful quantities}

In our approach to cosmology all physical quantities with dimensions (of
length, time, mass, etc.) are affected by the general \textquotedblleft
scaling\textquotedblright\ properties detailed by Eq. (\ref{eqn2.19}), with
the function $f(1+z)$ defined as in Eq. (\ref{eqn3.12}), where we use $a=2$ as
our preferred value for the inverse-power law generalization. In particular,
using the relations in Table I of our first paper, we can express all these
scaling factors as a function of $z$ and $\delta_{0}$, or as a function of the
\textquotedblleft cosmological time\textquotedblright\ $\delta$ and its
current value $\delta_{0}$, since we have:%

\begin{align}
1+z  &  =\sqrt{\frac{1-\delta_{0}^{2}}{1-\delta^{2}}}\label{eqn4.1}\\
f(1+z)  &  =\frac{4\delta_{0}^{2}(1+z)}{\left[  \delta_{0}(1+z)+\sqrt
{(1+z)^{2}-(1-\delta_{0}^{2})}\right]  ^{2}}=\frac{4\delta_{0}^{2}}{\left[
\delta_{0}+\left\vert \delta\right\vert \right]  ^{2}}\sqrt{\frac{1-\delta
^{2}}{1-\delta_{0}^{2}}}.\nonumber
\end{align}

These equations imply that when we observe the Universe at a certain fixed
value of the cosmological time (for example the current value $\delta_{0}$)
all dimensionful quantities and constants appear to scale with the redshift
parameter $z$, or with $\delta$ varying from $-1$ to $+1$. This will affect
also the fundamental constants of physics (with the exception of the speed of
light, as already discussed in our paper I).

For example, Planck's constant current value $h_{0}=6.626\times10^{-27}%
\ erg\ s$, would be perceived as scaling like the product of a mass times a
length, resulting in the following:%

\begin{equation}
h=h_{0}(1+z)f(1+z)=h_{0}\frac{4\delta_{0}^{2}(1+z)^{2}}{\left[  \delta
_{0}(1+z)+\sqrt{(1+z)^{2}-(1-\delta_{0}^{2})}\right]  ^{2}}=h_{0}\frac
{4\delta_{0}^{2}}{\left[  \delta_{0}+\left\vert \delta\right\vert \right]
^{2}}. \label{eqn4.2}%
\end{equation}

When we observe the Universe at a fixed cosmological time it is natural to
assume that all dimensionless quantities should also be considered fixed at
the particular values they have at that time. For example, the fine-structure
constant of electromagnetism is defined as the (dimensionless) quantity
$\alpha_{em}=e^{2}/\hbar c$ (with $\hbar=h/2\pi$) and has a current value of
$\alpha_{em_{0}}=7.297\times10^{-3}$. This definition of $\alpha_{em}$ implies
that the square of the elementary charge $e^{2}$ should scale as the Planck's
constant (since $c$ does not scale and $\alpha_{em}$ is assumed fixed, because
it's a dimensionless quantity). Therefore, the elementary charge would appear
to scale as:%

\begin{equation}
e=e_{0}\sqrt{(1+z)f(1+z)}=e_{0}\frac{2\delta_{0}(1+z)}{\left[  \delta
_{0}(1+z)+\sqrt{(1+z)^{2}-(1-\delta_{0}^{2})}\right]  }=e_{0}\frac{2\delta
_{0}}{\left[  \delta_{0}+\left\vert \delta\right\vert \right]  }.
\label{eqn4.3}%
\end{equation}

\subsection{\label{sect:dimensionless}Dimensionless vs. dimensionful
quantities}

Using the equations outlined in the previous sections we can compute the
scaling properties of all quantities with physical dimensions, as seen from an
observer at the current cosmological time $\delta_{0}$. However, the origin of
the variability of all these quantities should be found in the values of
dimensionless parameters and constants. In fact, we have shown above how the
scaling properties can be described in terms of a variable dimensionless
$\delta$ parameter.

An alternative, but equivalent way of describing these effects is to connect
them to a variable fine-structure constant $\alpha_{em}$. We recall that the
wavelength of emitted radiation, in a first-order approximation, can be
related to $\alpha_{em}$ as follows:%

\begin{equation}
\lambda\sim\frac{h}{mc\alpha_{em}^{2}}, \label{eqn4.4}%
\end{equation}
since the Rydberg constant\ for infinite nuclear mass is $R_{\infty}%
=\frac{m_{e}c\alpha_{em}^{2}}{2h}$ and the wavelength is inversely
proportional to $R_{\infty}$, as stated by the simple Balmer's formula. In Eq.
(\ref{eqn4.4}) $m$ can be considered to be the electron mass $m_{e}$ or the
reduced mass of the atomic system emitting the radiation.

If we assume that the cosmological redshift/blueshift is due to intrinsic
changes of $\alpha_{em}$, we have to consider the dimensionful constants $h$
and $m$ as fixed. In this case we can rewrite the first expression in Eq.
(\ref{eqn4.1}) as:%

\begin{equation}
1+z=\sqrt{\frac{1-\delta_{0}^{2}}{1-\delta^{2}}}=\frac{\lambda}{\lambda_{0}%
}=\frac{\alpha_{em_{0}}^{2}}{\alpha_{em}^{2}}, \label{eqn4.5}%
\end{equation}
thus obtaining how the fine-structure constant changes with $z$ or $\delta$:%

\begin{equation}
\alpha_{em}=\frac{\alpha_{em_{0}}}{\sqrt{1+z}}=\alpha_{em_{0}}\sqrt[4]%
{\frac{1-\delta^{2}}{1-\delta_{0}^{2}}}. \label{eqn4.6}%
\end{equation}

\begin{figure}
[ptb]
\begin{center}
\fbox{\ifcase\msipdfoutput
\includegraphics[
height=5.1422in,
width=6.3002in
]%
{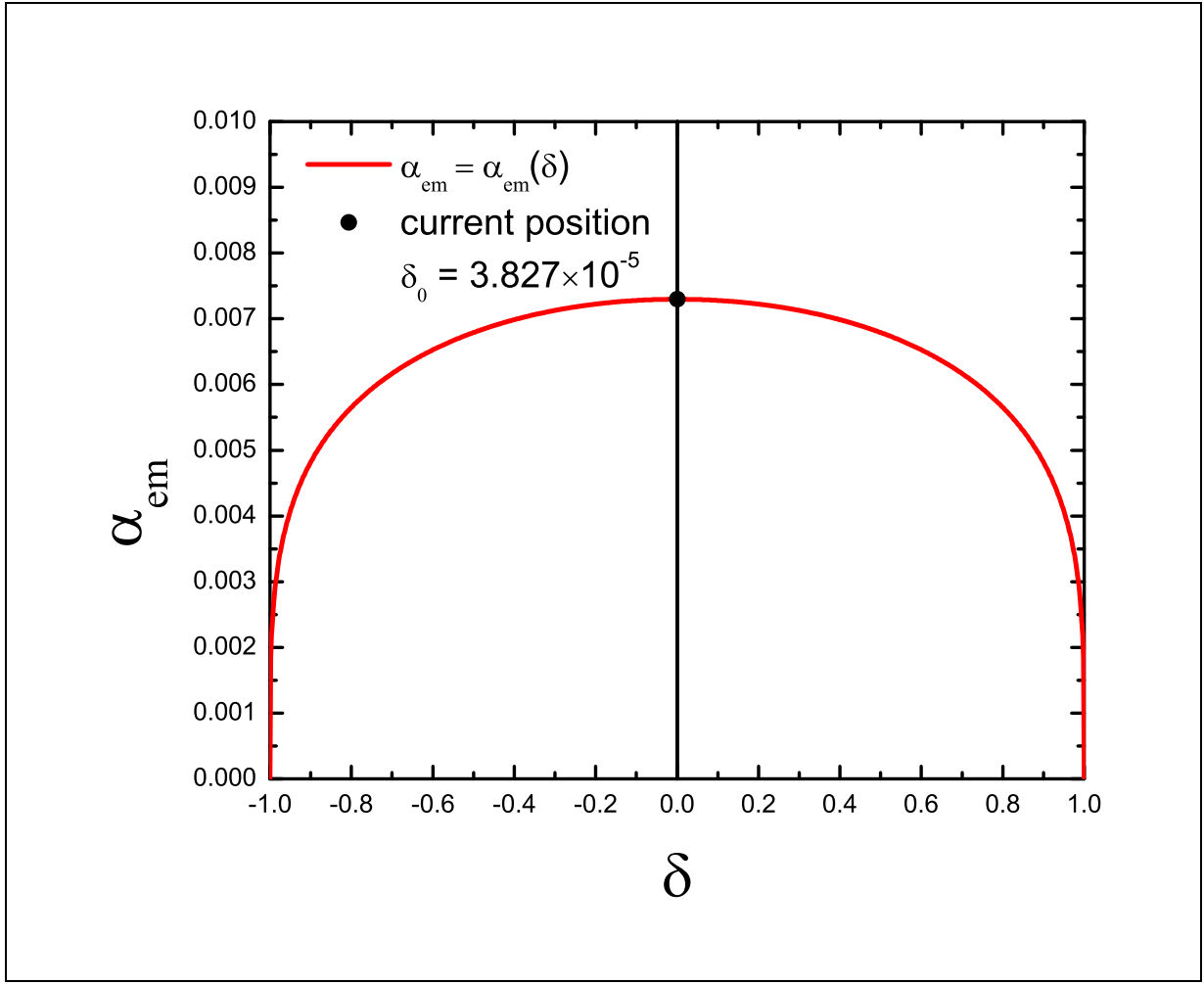}%
\else
\includegraphics[
height=5.1422in,
width=6.3002in
]%
{C:/swp35/Docs/KINEMATICAL2/graphics/figure5__5.pdf}%
\fi
}\caption[The dependance of $\alpha_{em}$ on our parameter $\delta$ is shown,
given the current values of these two quantities.]{The dependance of
$\alpha_{em}$ on our parameter $\delta$ is shown, given the current values of
these two quantities, $\alpha_{em_{0}}=7.297\times10^{-3}$ and $\delta
_{0}=3.827\times10^{-5}$ (red solid curve). Our current \textquotedblleft
position\textquotedblright\ in the plot is indicated by the black dot in the
figure.}%
\label{fig5}%
\end{center}
\end{figure}

This is illustrated in Fig. \ref{fig5}, where the dependence of $\alpha_{em}$
on our parameter $\delta$ is shown, given the current values of these two
quantities, $\alpha_{em_{0}}=7.297\times10^{-3}$ and $\delta_{0}%
=3.827\times10^{-5}$. The red continuous curve in the figure shows how the
fine-structure constant is zero for $\delta=-1$, it is increasing to its
maximum value $\alpha_{em}^{\max}=\alpha_{em_{0}}\sqrt[4]{1/(1-\delta_{0}%
^{2})}$ for $\delta=0$ and then decreasing to zero again for $\delta=+1$. Our
current \textquotedblleft position\textquotedblright\ in the plot is indicated
by the black dot in the figure. Given the very small value of $\delta_{0}$,
the current value $\alpha_{em_{0}}=7.297\times10^{-3}$ is basically the same
as the maximum, indicating that electromagnetic interactions are currently at
their strongest level.%

Therefore, if this hypothesis is correct, we can consider dimensionless
quantities such as $\delta$ and $\alpha_{em}$ as the fundamental physical
parameters directly connected to the evolution of the Universe. Their values
are changing in time, or actually they can be considered the fundamental
cosmological time. The values of the dimensionful constants and parameters are
merely a consequence of the current values of the dimensionless quantities,
which determine the standard units with which all quantities with physical
dimensions are measured.

In fact, A. Einstein was one of the first scientists to advocate for the
importance of dimensionless quantities over dimensionful ones, as shown in a
private correspondence with a former student of his (see discussion in Chapter
3 of Barrow's book \cite{Barrow1}). Einstein thought that constants with
physical dimensions are merely a product of the units of measure being used
and as such they do not possess a deep theoretical meaning. On the contrary,
dimensionless quantities constructed with standard dimensionful constants
(such as $\alpha_{em}=e^{2}/\hbar c$) are considered by Einstein to be the
only significant numbers in physics, whose value should be possible to explain
in terms of fundamental mathematical constants such as $\pi$ or $e$.

Our approach, described in this section, follows Einstein's suggestion of
dimensionless quantities as being the most fundamental ones, but we have shown
above that their values are also changing with the universal time. Therefore,
there is no need to explain a particular current value of $\alpha_{em}$ or
$\delta$, but it is sufficient to describe the evolution of these parameters
as it was done in this section.

We also remark that our description of the time variability of fundamental
constants is different from the standard approach to this subject (for reviews
see \cite{Barrow1}, \cite{Okun:2003wc}, \cite{Uzan:2002vq}). For example, in
recent claims of a (very small) time variability of the fine-structure
constant, as seen in interstellar absorption spectra \cite{Murphy:2000nr}, the
cosmological redshift is obviously factored out from the effect being studied,
thus resulting in a very small variation of $\alpha_{em}$ over cosmological
scales. On the contrary, in our approach the cosmological redshift is possibly
explained in terms of a (large) variation of the fine-structure constant, as
seen in Fig. \ref{fig5}, therefore resulting in a totally different phenomenology.

It is beyond the scope of this work to extend this analysis to all other
fundamental constants in nature (dimensionless or not). We simply point out
that if our hypothesis on the variability of physical constant is correct,
this would call for a revision of the theory of fundamental interactions, such
as quantum electrodynamics or others, in view of variable coupling constants
and interaction strengths.

\subsection{\label{sect:observations_experiments}Astrophysical observations
and kinematical conformal cosmology}

We have seen that in our approach to cosmology only experiments and
observations based on atomic properties and electromagnetic phenomenology are
affected by our interpretation of the redshift. Gravity in itself is not
necessarily affected, as described by the Newton-Einstein paradigm; on the
contrary there is a possible change in electromagnetic physics (and perhaps
also in strong/weak interactions) over cosmological times, which affects our
observations and perspective of the gravitational motion.

In other words, the description of the Universe can still be done with a
standard Newton-Einstein paradigm, assuming invariable space-time units, if we
measure all quantities with these fixed units, as in the case when we use
\textquotedblleft ranging\textquotedblright\ techniques, i.e., studying the
time of flight of light signals to determine the positions of celestial bodies.

On the contrary, certain other astrophysical observations (such as those based
on spectroscopy or similar) need to include the stretching of space-time and
the change of related units. In this case we need Conformal Gravity as an
enhancement of General Relativity, to include this \textquotedblleft
stretching\textquotedblright\ of the units and our kinematical approach might
help to explain the shortcomings of the standard theory.

In order to test our model, a conclusive experiment would be an enhanced
measurement of the \textquotedblleft Pioneer anomaly\textquotedblright%
\ effect, possibly realized through a dedicated mission of a spaceship in the
outer Solar System, as it has already been proposed (\cite{Anderson:2002yc},
\cite{Nieto:2004np}, \cite{Dittus:2005re}). This seems to be the only
practicable type of experiment in which an electromagnetic signal of well
known wavelength can be transmitted unperturbed over a considerable temporal
interval. In this way, it could be compared to a similar signal produced at a
later time to check for an intrinsic wavelength shift, due to the conformal
stretching of the space-time. The direct detection of such a
wavelength/frequency shift of an electromagnetic signal with time would be a
clear signature of Conformal Gravity acting as described by our kinematical approach.

\section{\label{sect:conclusions}Conclusions}

We have introduced experimental evidence in support of our kinematical
conformal cosmology and determined the values of its fundamental parameters.
In particular, we have focused our analysis on reproducing the Hubble plots
for type Ia Supernovae, with the same level of accuracy obtained by standard cosmology.

To achieve this goal we critically reconsidered all the standard distances
commonly used in cosmology and we added a new scaling property for the masses
(or the energies) as a function of the redshift parameter. Our new expression
for the distance modulus as a function of redshift can effectively fit the
\textquotedblleft gold-silver\textquotedblright\ SNe data with the required
accuracy and also yield a current-time value for our fundamental parameter
$\delta_{0}$, which is small and positive as expected.

Since type Ia Supernovae, or other astrophysical candles, are distant
cosmological objects, our second point of focus was to consider more local
effects due to our kinematical conformal cosmology, which might be more
suitable for the determination of the parameters. In particular, a local
blueshift region was expected, given the estimated values of the quantities in
our model, and is possibly evidenced by the recently discovered Pioneer anomaly.

We have seen how our model can account for this effect and can be used to
estimate our second fundamental parameter $\gamma_{0}$, which together with
$\delta_{0}$ will determine all the other quantities in our model. More
precise evaluations of these parameters are certainly needed and should come
from an extended analysis of the Pioneer data or through a dedicated future
spacecraft mission, which has already been proposed.

We argued that a direct detection of a frequency/wavelength shift in
electromagnetic radiation, traveling over distances comparable to the size of
our Solar System, could be explained only in terms of a conformal space-time
stretching and would be the best evidence in support of our kinematical
approach. If the Pioneer anomaly, or similar phenomena, will prove in the next
few years to be positive indications of these effects, this might also signal
a possible time-variation of dimensionless fundamental quantities, such as the
fine-structure constant and our cosmological time $\delta$. This would
constitute an important step towards a deeper understanding of the role and
values of all the fundamental physical constants and also impact our current
understanding of the fundamental interactions.

\section{\label{sect:appendix}Appendix I - K-corrections in luminosity
measurements}

In this appendix we will briefly review the theory of K-corrections used in
luminosity measurements of standard candles (especially type Ia SNe) and check
if our new cosmology requires any changes in the definition of such
corrections. We recall that a luminosity distance vs. magnitude equation, such
as Eq. (\ref{eqn2.17}) used in standard cosmology or our revised formula
(\ref{eqn3.13}), usually relate the theoretical expression of $d_{L}$ to the
\textquotedblleft bolometric\textquotedblright\ apparent and absolute
magnitudes of the light source, i.e., considering radiation emitted over the
whole wavelength spectrum.

On the contrary, CCD detectors or other photometric devices used in astronomy
usually observe radiation within certain wavelength bands (such as those
related to the $U$, $B$, $V$, $R$, $I$ filters mentioned in Sect.
\ref{sect:supernovae}) and this information needs to be converted into
bolometric magnitudes before it can be used in the luminosity distance
equations. \textquotedblleft K-corrections\textquotedblright\ are introduced
for this purpose. Moreover, due to the large redshifts of some of the SNe
being observed, photometry in a highly redshifted\ filter (such as in the
$R$-band) has sometimes to be compared to nearby photometry of a reference
Supernova in a different filter (such as $B$, $V$ or other), thus involving
also conversions between different filters, in addition to the standard
correction from one band to bolometric magnitudes.

The theory of K-corrections was originally introduced by Humason, Mayall and
Sandage \cite{1956AJ.....61...97H} and later reviewed by Oke and Sandage
\cite{1968ApJ...154...21O}\ in 1968. It was later adapted to the modern case
of type Ia SNe spectra in a series of papers (see for example
\cite{Kim:1995qj}, \cite{Kim:1996an}, \cite{Nugent:2002si}). In this Appendix
we will follow the notation used in these modern reviews.

The standard K-correction connects the apparent magnitude $m_{x}$ in some
\textquotedblleft$x$\textquotedblright\ filter band of a light source at
redshift $z$, to the distance modulus $\mu(z)$ according to the following
equation \cite{Kim:1995qj}:%

\begin{align}
\mu(z)  &  =m_{x}(z,t_{obs})-M_{x}(z=0,t_{rest})-K_{x}(z,t_{rest}%
)\label{eqn6.1}\\
K_{x}(z,t_{rest})  &  =2.5\log_{10}(1+z)+2.5\log_{10}\left\{  \frac{%
{\textstyle\int}
F(\lambda)S_{x}(\lambda)d\lambda}{\int F\left[  \lambda/(1+z)\right]
S_{x}(\lambda)d\lambda}\right\}  ,\nonumber
\end{align}
where $M_{x}$ is the absolute $x$ magnitude and the appropriate correction
$K_{x}$ is detailed in the second line of the equation, with $F(\lambda)$
being the spectral energy distribution \textquotedblleft at the
source\textquotedblright\ and $S_{x}(\lambda)$ the filter transmission. The
time variables are connected by the standard time dilation equation
$t_{obs}=t_{rest}(1+z)$, where $t_{rest}$ is the time in the Supernova rest
frame and $t_{obs}$ is the time in the observer frame. These times correspond
respectively to our times $t_{z}$ and $t_{0}$ as discussed in section
\ref{sect:luminosity_distance}, related in the same way, but with a different
interpretation. We will omit the time dependence in the following since it's
not essential for our discussion.

Following our revision of the luminosity distance, detailed in Sect.
\ref{sect:luminosity_distance}, we will recompute in our new notation the
$K_{x}$ term and compare it to the standard expression in Eq. (\ref{eqn6.1}).
Since $\mu(z)=m(z)-M(z=0)$, where $m$ and $M$ are bolometric magnitudes, the
simple K-correction is actually defined as:%

\begin{align}
K_{x}(z)  &  =\left[  m_{x}(z)-M_{x}(z=0)\right]  -\left[  m(z)-M(z=0)\right]
=\label{eqn6.2}\\
&  =-2.5\log_{10}\left[  l_{x}(z)/l_{x}(z=0)\right]  +2.5\log_{10}\left[
l_{bol}(z)/l_{bol}(z=0)\right]  =\nonumber\\
&  =-2.5\log_{10}\left[  \frac{%
{\textstyle\int}
\frac{dL_{z}}{d\lambda}(\lambda_{z})S_{x}(\lambda_{z})d\lambda_{z}}{4\pi
d_{L}^{2}}\frac{4\pi d_{ref}^{2}}{\int\frac{dL_{0}}{d\lambda}\left(
\lambda_{0}\right)  S_{x}(\lambda_{0})d\lambda_{0}}\frac{L_{0}}{4\pi
d_{ref}^{2}}\frac{4\pi d_{L}^{2}}{L_{z}}\right]  =\nonumber\\
&  =2.5\log_{10}\left(  \frac{L_{z}}{L_{0}}\right)  +2.5\log_{10}\left[
\frac{\int F_{0}\left(  \lambda_{0}\right)  S_{x}(\lambda_{0})d\lambda_{0}}{%
{\textstyle\int}
F_{z}(\lambda_{z})S_{x}(\lambda_{z})d\lambda_{z}}\right]  =\nonumber\\
&  =2.5\log_{10}\left[  \frac{f(1+z)}{(1+z)}\right]  +2.5\log_{10}\left[
\frac{\int F_{0}\left(  \lambda_{0}\right)  S_{x}(\lambda_{0})d\lambda_{0}%
}{\frac{f(1+z)}{(1+z)^{2}}%
{\textstyle\int}
F_{0}(\lambda_{0})S_{x}(\lambda_{z})d\lambda_{z}}\right]  =\nonumber\\
&  =2.5\log_{10}\left[  \frac{f(1+z)}{(1+z)}\right]  +2.5\log_{10}\left[
\frac{(1+z)^{2}}{f(1+z)}\right]  +2.5\log_{10}\left[  \frac{\int F_{0}\left(
\lambda_{0}\right)  S_{x}(\lambda_{0})d\lambda_{0}}{%
{\textstyle\int}
F_{0}\left[  \lambda_{z}/(1+z)\right]  S_{x}(\lambda_{z})d\lambda_{z}}\right]
=\nonumber\\
&  =2.5\log_{10}(1+z)+2.5\log_{10}\left[  \frac{\int F_{0}\left(
\lambda\right)  S_{x}(\lambda)d\lambda}{%
{\textstyle\int}
F_{0}\left[  \lambda/(1+z)\right]  S_{x}(\lambda)d\lambda}\right]  .\nonumber
\end{align}

In this derivation we used our new notation for all the functions and
variables involved, including the connection between wavelengths $\lambda
_{z}=\lambda_{0}(1+z)$, the spectral distribution \textquotedblleft at the
source\textquotedblright\ $F_{0}\left(  \lambda_{0}\right)  \equiv\frac
{dL_{0}}{d\lambda}\left(  \lambda_{0}\right)  $ and the spectral distribution
with the source placed at redshift $z$ and observed from the origin,
$F_{z}\left(  \lambda_{z}\right)  \equiv\frac{dL_{z}}{d\lambda}\left(
\lambda_{z}\right)  $. We also used Eqs. (\ref{eqn2.21}), (\ref{eqn2.23}) in
the chain of derivation and renamed in the last line the wavelengths
$\lambda_{z}$, $\lambda_{0}$ simply as $\lambda$, to obtain exactly the
standard result of Eq. (\ref{eqn6.1}).

Therefore, the simple K-corrections computed according to the standard theory
are unchanged in our kinematical conformal cosmology, although, by checking
the derivation of Eq. (\ref{eqn6.2}) and comparing it to the same derivation
of the standard theory, we observe that the final two terms originate in a
slightly different way (see for example how the $f(1+z)$ function cancels out
in our case) and do not have the same meaning as those in the standard theory
\cite{1968ApJ...154...21O}.

When two different filters are used in the observations, for example the
high-redshift photometry is observed with a \textquotedblleft$y$%
\textquotedblright\ filter and related to the nearby reference photometry in
the \textquotedblleft$x$\textquotedblright\ filter, a new $K_{xy}$ correction
is used, defined as \cite{Kim:1995qj}:%

\begin{align}
\mu(z)  &  =m_{y}(z,t_{obs})-M_{x}(z=0,t_{rest})-K_{xy}(z,t_{rest}%
)\label{eqn6.3}\\
K_{xy}(z,t_{rest})  &  =2.5\log_{10}(1+z)+2.5\log_{10}\left\{  \frac{%
{\textstyle\int}
F(\lambda)S_{x}(\lambda)d\lambda}{\int F\left[  \lambda/(1+z)\right]
S_{y}(\lambda)d\lambda}\right\}  -2.5\log_{10}\left\{  \frac{%
{\textstyle\int}
Z(\lambda)S_{x}(\lambda)d\lambda}{\int Z\left(  \lambda\right)  S_{y}%
(\lambda)d\lambda}\right\}  .\nonumber
\end{align}

The meaning of the terms and variables is similar to that of Eq.
(\ref{eqn6.1}) and the main difference is the addition of a third term in the
K-correction, which accounts for the differences in the zero points of the two
filters ($Z\left(  \lambda\right)  $ is an idealized stellar spectral energy
distribution at $z=0$ for which $U=B=V=R=I=0$ in the photometric system being
used \cite{Kim:1995qj}). The previous expression reduces to the simple K-term
for the case when the two filters are the same and the third term added in Eq.
(\ref{eqn6.3}) is a mere technical correction due to the different zero points
used, thus unaffected by our new approach.

In conclusion, our new approach does not practically change the computation of
K-corrections as done in the standard theory, therefore it was correct to use
the experimental values of the distance moduli of the \textquotedblleft
gold-silver\textquotedblright\ type Ia SNe data \cite{Riess:2006fw} in our
analysis in Sect. \ref{sect:supernovae}.

\bibliographystyle{apsrev}
\bibliography{CMB,CONFORMAL,CONSTANTS,COSMOBOOKS,HSTKEY,MOND,OTHERCMB,PIONEER,PK,SDSS,SUPERNOVAE}

\end{document}